\documentclass[aip,
nofootinbib,
rsi,%
reprint,
twocolumn,
longbibliography,
author-numerical%
]{revtex4-2}

\usepackage[T2A]{fontenc}
\usepackage{graphicx}
\usepackage{textcomp}
\usepackage{natbib}

\usepackage{amsfonts}
\usepackage{amssymb}
\usepackage{amsmath}
\usepackage[usenames]{color}

\usepackage{setspace}
\usepackage{changes}

\newcommand{\tc}{\tilde{c}}

\newcommand{\teta}{\tilde{\eta}}

\newcommand{\tJ}{\tilde{J}}

\newcommand{\tZ}{\tilde{Z}}
\newcommand{\tD}{\tilde{D}}

\newcommand{\tl}{\tilde{l}}

\newcommand{\expo}{{{\rm e}^{\teta_0}}}
\newcommand{\expot}{{{\rm e}^{2\teta_0}}}

\newcommand{\Rinf}{R_{\infty}}
\newcommand{\Rpol}{R_{pol}}

\newcommand{\cref }{c_h^{in}}

\newcommand{\tom  }{\tilde{\omega}}

\newcommand{\jlim}{j_{\lim}}

\newcommand{\Cdl}{C_{dl}}

\newcommand{\lam }{\lambda}

\newcommand{\veps}{\varepsilon_*}

\newcommand{\sion}{\sigma_p}

\newcommand{\Dox}{D_{ox}}

\newcommand{\lcat}{l_t}

\newcommand{\ri}{{\rm i}}

\newcommand{\sqa}{\ri\tom/\tD_b}

\newcommand{\lexp}[1]{\exp\left(#1\right)}
\newcommand{\lnl }[1]{\ln \left(#1\right)}

\renewcommand{\Re}[1]{\operatorname{Re}\left(#1\right)}

\newcommand{\etal}{et al.{ }}

\renewcommand{\Re}[1]{\operatorname{Re}\left(#1\right)}
\renewcommand{\Im}[1]{\operatorname{Im}\left(#1\right)}

\begin{document}

\sf

\title{A kernel for PEM fuel cell distribution of relaxation times}



\author{Andrei Kulikovsky}
\thanks{ECS Active member}
\email{A.Kulikovsky@fz-juelich.de}

\affiliation{Forschungszentrum J\"ulich GmbH           \\
Theory and Computation of Energy Materials (IEK--13)   \\
Institute of Energy and Climate Research,              \\
D--52425 J\"ulich, Germany
}

\altaffiliation[Also at: ]{Lomonosov Moscow State University,
                Research Computing Center, 119991 Moscow, Russia}

\date{\today}

\begin{abstract}
Impedance of all oxygen transport processes in PEM fuel cell
has negative real part in some frequency domain. 
A model function (kernel) for calculation of distribution of relaxation times (DRT)
of a PEM fuel cell is suggested. The kernel is designed for capturing 
impedance with negative real part  and it stems from the 
equation for impedance of oxygen transport through the gas--diffusion transport layer
(doi:10.1149/2.0911509jes). Using recent analytical solution 
for the cell impedance it is shown that DRT calculated with the novel $K_2$ kernel
correctly captures the GDL transport peak, while the classic DRT based 
on the $RC$--circuit (Debye) kernel misses this peak.  
Employing $K_2$ kernel, analysis of DRT spectra of a real PEMFC is performed. 
The leftmost on the frequency scale DRT peak represents oxygen transport in the channel, 
and the rightmost peak is due to proton transport in the cathode catalyst layer. 
The second, third and fourth peaks exhibit oxygen transport in the GDL, faradaic 
reactions on the cathode side, and oxygen transport in the catalyst layer, respectively.
\end{abstract}


\maketitle

\section{Introduction}

Electrochemical impedance spectroscopy provides unique opportunity 
for testing and characterization of PEM fuel cells without interruption of 
current production mode\cite{Lasia_book_14}. A classic approach to interpretation 
of EIS data is construction of equivalent electric circuit having impedance 
spectrum close to the measured one. However, a more attractive 
option provides the distribution of relaxation times (DRT) technique. 

In the context of PEM fuel cell studies, the idea of DRT 
can be explained as following. To a first approximation, PEM fuel cell
impedance $Z$ can be modeled as impedance of a parallel $RC$--circuit
\begin{equation}
   Z = \dfrac{R}{1 + \ri\omega R C}.
   \label{eq:ZRC}
\end{equation}
where $\omega$ is the angular frequency of applied AC signal.
This approximation corresponds to the cell with ideal (fast) transport
of reactants in all transport medias\cite{Kulikovsky_17b}.
In that case, $R$ describes Tafel resistivity 
of the oxygen reduction reaction (ORR) and $C$ represents the superficial 
double layer capacitance of the electrode.

In general, to calculate transport contributions to cell impedance 
one has to develop a transport model for 
the ORR reactants. However, if information on transport resistivities and
characteristic frequencies suffice, DRT provides another option.
Denoting $RC = \tau$, multiplying the right side of Eq.\eqref{eq:ZRC} 
by nonnegative function $\gamma(\tau)$ and integrating over $\tau$, we get
\begin{equation}
   Z(\omega) = \Rpol \int_0^{\infty}\dfrac{\gamma(\tau)d\tau}{1 + \ri\omega\tau}
   \label{eq:drt0}
\end{equation} 
where $\Rpol$ is the total polarization resistivity of the cell and
the function $\gamma$ is the DRT of impedance $Z$. Mathematically, Eq.\eqref{eq:drt0} 
means expansion of $Z(\omega)$ into infinite sum of $RC$--impedances,
with the resistivity of each elementary $RC$--circuit being $R_{pol}\gamma\,d\tau$. 
The function
\begin{equation}
   K_{RC}(\omega, \tau) =  \dfrac{1}{1 + \ri\omega\tau}
   \label{eq:Krc}
\end{equation}
under integral in Eq.\eqref{eq:drt0} is usually called 
``Debye model''\cite{BarsMac_book_18}.
Eq.\eqref{eq:drt0} can also be considered as integral transform 
of $\gamma(\tau)$, which justifies the term ``$RC$--kernel'' for Eq.\eqref{eq:Krc}.  

Quite evidently, DRT of a single parallel $RC$--circuit 
is Dirac delta--function $\gamma = \delta(\tau - \tau_*)$
positioned at $\tau_* = RC$. This example illustrates the main feature 
of DRT: it converts any $RC$--like impedance into a single, more or less 
smeared  on the $\tau$--scale $\delta$--like peak.

All transport processes in a fuel cell eventually are linked to the double 
layer capacitance in the catalyst layer; thus, it is usually assumed that 
the impedance of every process is not far from impedance 
of a parallel $RC$--circuit. That means that the DRT of a PEMFC  
is expected to consist of several delta--like peaks. Since the regular frequency 
$f = 1 / (2\pi\tau)$, it is convenient to plot $\gamma(f)$ instead 
of $\gamma(\tau)$. Position of each peak on the frequency scale marks a 
characteristic frequency of the respective transport process, and 
\begin{equation}
   R_{n} = \Rpol\int_{\tau_n}^{\tau_{n+1}}\gamma(\tau)\,d\tau  
     \label{eq:Rn}
\end{equation}
gives the contribution of process resistivity in the total cell polarization 
resistivity $R_{pol}$. Here $\tau_n$ and $\tau_{n+1}$ are the peak boundaries
on the $\tau$--scale. 

Fuel cell impedance is usually measured on equidistant in log--scale frequency 
mesh $\{f_n, n=1,\ldots,N\}$ with
$\ln(f_{n+1}) - \ln(f_n)$ being independent of $n$. From numerical perspective 
it is beneficial to deal with the function $G(\tau)$ satisfying to
\begin{equation}
   Z(\omega) = \Rinf + R_{pol}\int_{-\infty}^{\infty}\dfrac{G(\tau)\,d\ln(\tau)}{1 + \ri\omega\tau},
   \label{eq:drtG}
\end{equation}
where the term $\Rinf$ is added to describe pure ohmic (high--frequency) 
fuel cell resistivity.
Clearly, $\gamma = G / \tau$ and 
Eq.\eqref{eq:Rn} in terms of $G$ takes the form
\begin{equation}
   R_n = \Rpol\int_{\tau_n}^{\tau_{n+1}}G(\tau)\,d\ln\tau \\
       = 2\pi \Rpol \int_{f_{n}}^{f_{n+1}} G(f)\,d\ln(f),
   \quad f_n < f_{n+1}
   \label{eq:RnG}
\end{equation}
where the frequencies $f_n = 1/(2\pi \tau_{n+1})$ 
and $f_{n+1} = 1 / (2\pi \tau_n)$ mark the peak boundaries.  
Setting in Eq.\eqref{eq:drtG} $\omega=0$ and taking into account that 
$Z(0) - \Rinf = \Rpol$, we see that $G$ obeys to normalization condition
\begin{equation}
   \int_{-\infty}^{\infty} G(\tau)\,d\ln\tau = 2\pi\int_{-\infty}^{\infty} G(f)\,d\ln f = 1
   \label{eq:normG}
\end{equation}
In the following, Eq.\eqref{eq:drtG} will be discussed, as $G$ is usually
used instead of $\gamma$ is practical calculations.


DRT technique, Eq.\eqref{eq:drt0}, was invented by Fuoss and Kirkwood in 1941\cite{Fuoss_41}
in the context of polymer materials impedance   
and brought to the fuel cell community seemingly by Schlichlein \etal\cite{Schlichlein_02}. 
Since 2002, a lot of works from the group of Ivers--Tiff{\'e}e have been devoted to deciphering of solid oxide 
fuel cell spectra by means of DRT (see a review\cite{Tiffee_17}). 
Analysis of PEMFC impedance spectra using DRT is 
a relatively new field\cite{Heinzmann_18,Tsur_21,Kulikovsky_21f,Minggao_21}.
Heinzmann, Weber and Ivers--Tiff{\'e}e\cite{Heinzmann_18} measured impedance 
spectra of a small (1 cm$^2$) laboratory 
PEMFC and studied DRT peaks behavior depending on 
cell temperature, relative humidity (RH), oxygen concentration and current density.
They obtained DRT with up to five peaks; 
the leftmost on the frequency scale peak P1 was attributed 
to oxygen diffusion 
in the gas--diffusion and cathode catalyst layers. Cohen, Gelman and Tsur\cite{Tsur_21}
performed impedance measurements of a 5--cm$^2$ cell varying temperature,
RH and current density. The calculated DRT consisted of four peaks, attributed 
 (in ascending frequencies) to (1) oxygen transport in the GDL / CCL, (2) ORR,  
(3) proton transport in the CCL, and (4) proton transport in membrane. 
Note that Heinzmann \etal and Cohen \etal 
used different codes for DRT calculation.  Wang \etal\cite{Minggao_21} measured impedance
spectra of application--relevant 25--cm$^2$ PEMFC and obtained a three--peak DRT; 
the lowest frequency peak was attributed to oxygen diffusion processes 
in the cell. In our recent work\cite{Kulikovsky_21f} DRT spectra of a low--Pt PEMFC
have been reported; we attributed the low--frequency peak to oxygen transport 
in the GDL and, possibly, in the channel.  

In PEMFCs, the supplied oxygen (air) is transported through the four quite different medias:
channel, GDL, open pores of the CCL, and finally through Nafion film covering 
Pt/C agglomerates. One, therefore, could expect four corresponding peaks in the DRT 
spectra. However, in\cite{Heinzmann_18,Tsur_21,Minggao_21}, a single oxygen transport 
peak has been reported. There are two options to explain this result: either some 
of the oxygen transport peaks overlap with each other (or with the ORR peak) and 
DRT is not able to separate them, or the codes used were unable to resolve all the transport 
processes. It is important to note that the code for DRT calculation 
of Wan \etal\cite{Wan_15} used 
in\cite{Heinzmann_18,Minggao_21}, the ISGP code\cite{Tsur_11} used in\cite{Tsur_21},
and our code employing Tikhonov regularization 
in combination with NNLS solver\cite{Kulikovsky_20b,Kulikovsky_21b} 
are based on the $RC$--kernel, Eq.\eqref{eq:drtG}.  

The real part of $RC$--circuit impedance, Eq.\eqref{eq:ZRC}, 
is positive and imaginary part is negative. This imposes 
limits on functions $Z$ which could be represented by Eq.\eqref{eq:drtG}.
For example, impedance of an inductive loop cannot be expanded in infinite 
series of $RC$--impedances, as imaginary part of inductive impedance is 
positive. Quite similarly, the impedance $Z$ having negative real part
in some frequency domain also cannot be represented by Eq.\eqref{eq:drtG}.    

Below, we show that the DRT calculated with $RC$--kernel could completely miss 
some of the transport peaks in PEMFCs spectra. An alternative $K_2$ kernel better capturing 
oxygen transport processes in the cell cathode is suggested. The kernel 
is illustrated by calculation of DRT of the recent analytical PEM fuel cell 
impedance spectrum\cite{Kulikovsky_21h}. Finally, we show that the new kernel 
well separates the channel, GDL and ORR peaks in the DRT spectra 
of a standard Pt/C--based PEM fuel cell.

\section{Model: $K_2$ kernel}

Below, the following dimensionless variables will be used
\begin{multline}
                          \tc = \dfrac{c}{\cref},
                          \quad \tJ = \dfrac{J}{i_*\lcat},
                          \quad \teta = \dfrac{\eta}{b}, \quad
                          \tD_b = \dfrac{4 F D_b \cref}{i_*\lcat^2}, \\
                          \quad \tl_b = \dfrac{l_b}{\lcat},
                          \quad \tom = \omega t_*,
                          \quad \tZ = \dfrac{Z i_* \lcat}{b}.
   \label{eq:dless}
\end{multline}
Here,
$t_*$ is the characteristic time of double layer charging
\begin{equation}
   t_* = \dfrac{\Cdl b}{i_*},
   \label{eq:tast}
\end{equation}
$c$ is the oxygen concentration,
$\cref$ is the reference oxygen concentration,
$\Cdl$ is the volumetric double layer capacitance,  
$J$ is the mean current density in the cell,
$\eta$ is the ORR overpotential, positive by convention,
$b$ is the ORR Tafel slope, 
$i_*$ is the volumetric ORR exchange current density,
$\lcat$ is the cathode catalyst layer (CCL) thickness,
$D_b$ is the oxygen diffusion coefficient in the GDL,
$l_b$ is the GDL thickness. 

Analytical GDL impedance $\tZ_{gdl}$ has been derived in\cite{Kulikovsky_15g}. 
For the cell current densities well below the limiting current density 
due to oxygen transport in the GDL, $\tZ_{gdl}$ has the form
\begin{equation}
   \tZ_{gdl} = \dfrac{\tanh\left(\mu\tl_b\sqrt{\sqa}\right)}
                     {\mu \sqrt{\ri\tom\tD_b}\left(1 + \ri\tom/\tJ\right)}
   \label{eq:tZgdl}
\end{equation} 
where $\mu$ is the constant parameter
\begin{equation}
   \mu = \sqrt{\dfrac{4 F \cref}{\Cdl b}},
   \label{eq:mu}
\end{equation} 
Eq.\eqref{eq:tZgdl} is a Warburg finite--length impedance 
divided by the factor $(1 + \ri\tom/\tJ)$. The factor describes  the effect 
of double layer charging by the cell current density $\tJ$ transported 
in the form of oxygen flux through the GDL to the attached CCL\cite{Kulikovsky_15g}.
In his classic work\cite{Warburg_1899}, Warburg used static polarization curve 
to derive the boundary condition for calculation of transport impedance of a semi--infinite 
electrode (Eq.(4) of Ref.\cite{Warburg_1899}). 
Later, Warburg model has been extended for the case of finite--length 
transport layer, using the same boundary condition 
on the electrode side (see Ref.\cite{Lasia_book_14}, page 104). 
However, account of capacitive term in the electrode charge conservation 
equation changes the boundary condition for the oxygen transport equation\cite{Kulikovsky_15g},
leading to the additional factor $(1 + \ri\tom/\tJ)$ in denominator of Eq.\eqref{eq:tZgdl}.  

\begin{figure}
\begin{center}
\includegraphics[scale=0.45]{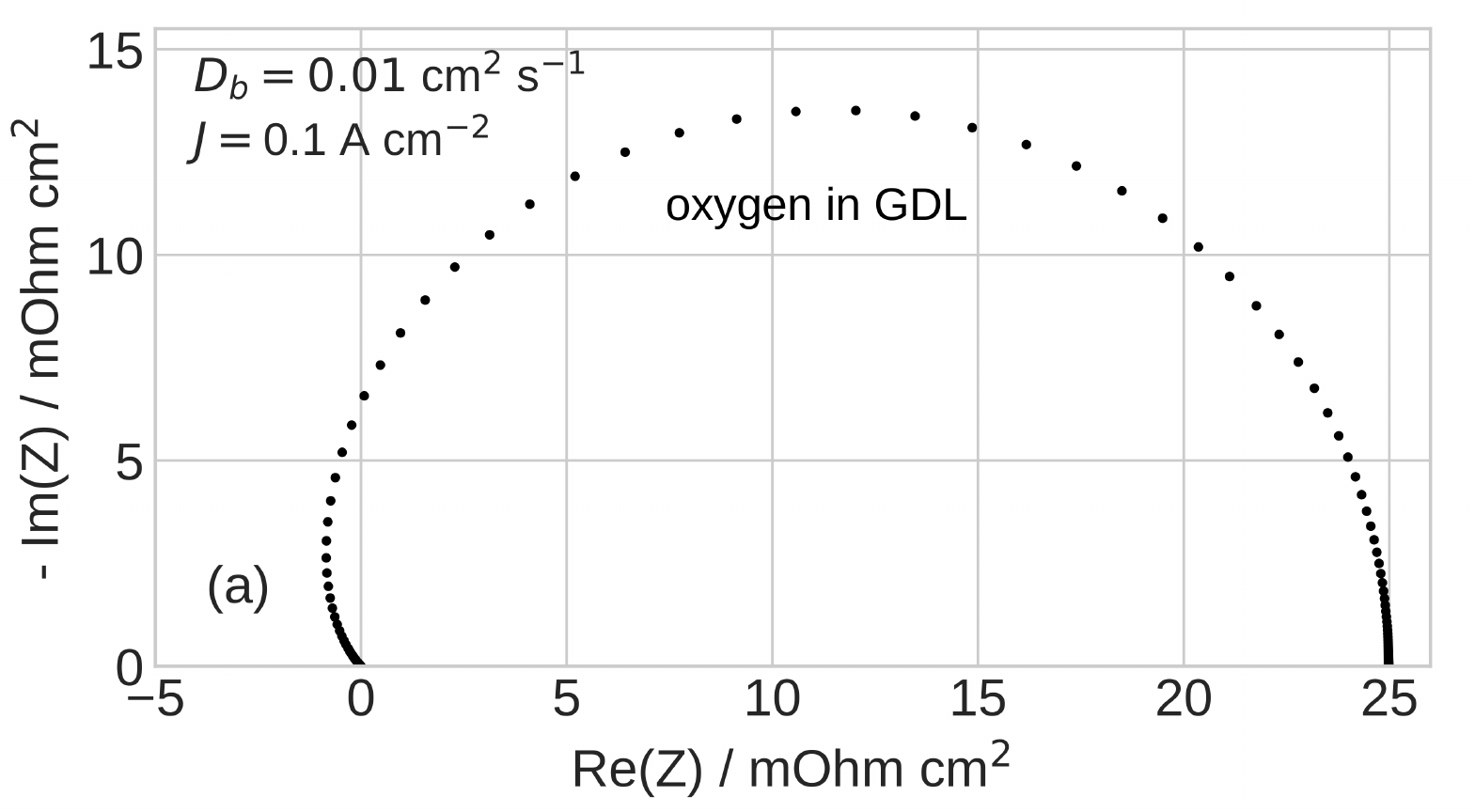}
\includegraphics[scale=0.45]{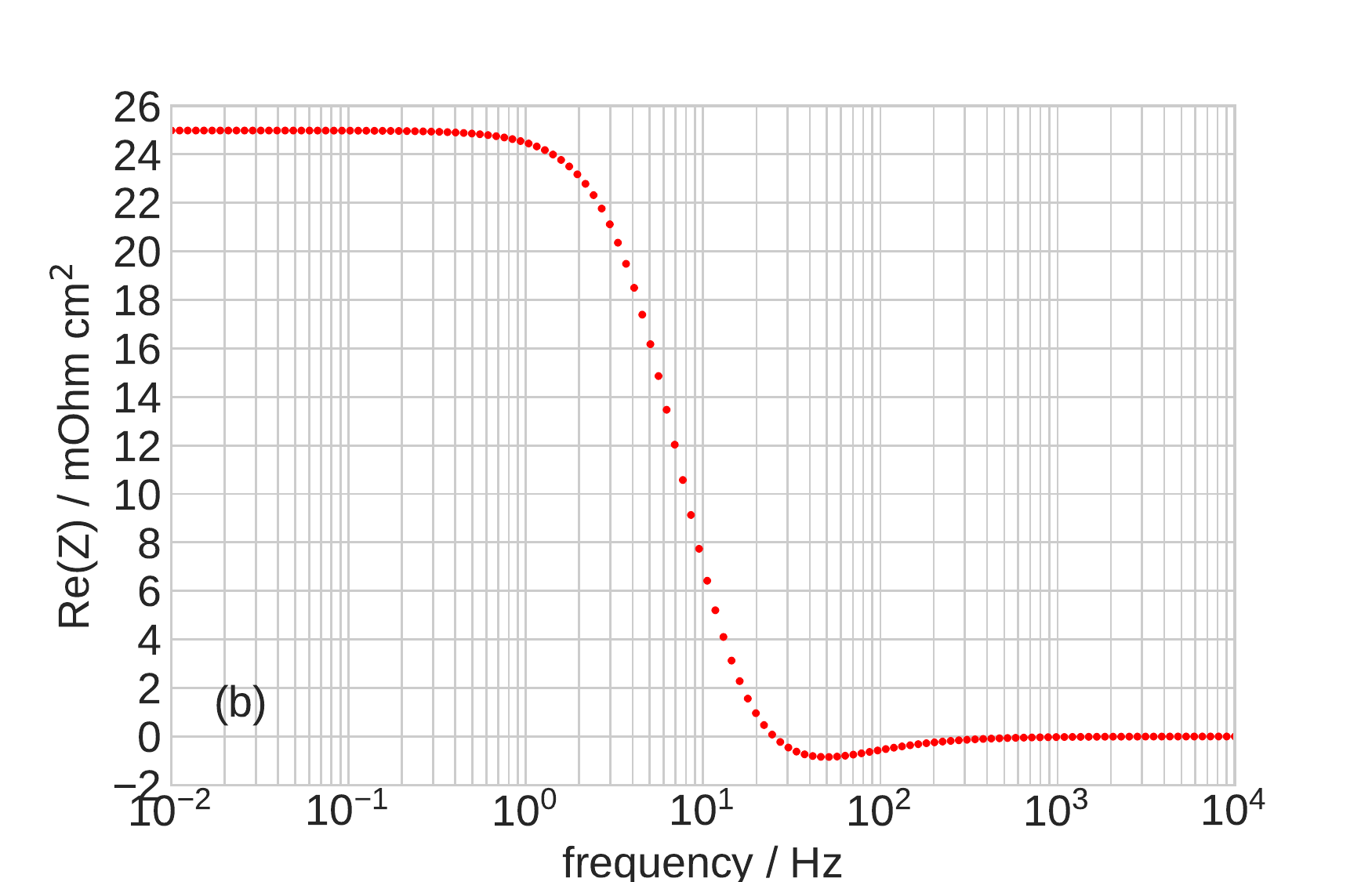}
\caption{(a) Nyquist spectrum of the gas--diffusion layer impedance, Eq.\eqref{eq:tZgdl}.
   (b) Frequency dependence of the real part of impedance in (a).  
   Parameters for calculations are listed in Table~\ref{tab:parms}.
  }
\label{fig:Zgdl}
\end{center}
\end{figure}

Nyquist plot of impedance \eqref{eq:tZgdl} is shown in Figure~\ref{fig:Zgdl}a;
frequency dependence of $\Re{\tZ_{gdl}}$ is depicted in Figure~\ref{fig:Zgdl}b.
As can be seen, between 20 and 200 Hz 
the real part of $Z_{gdl}$ is essentially negative, and 
at higher frequencies $\Re{\tZ_{gdl}}$ tends to zero.  

%

\begin{figure}
\begin{center}
\includegraphics[scale=0.45]{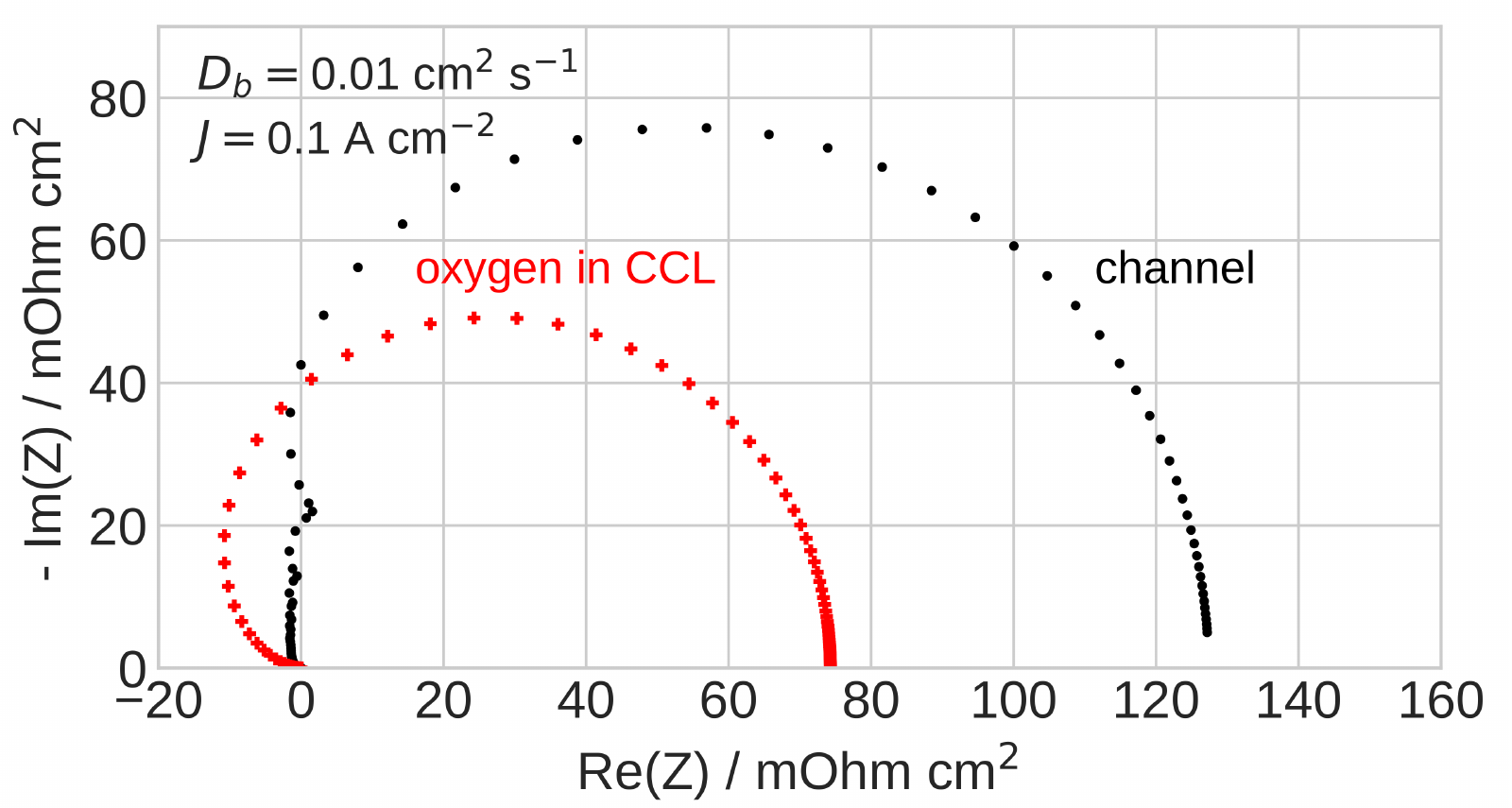}
\caption{Nyquist spectra of the channel impedance, Eq.\eqref{eq:tZchan} and
   of impedance of oxygen transport in the CCL\cite{Kulikovsky_17b}.
}
\label{fig:Zox}
\end{center}
\end{figure}

Impedance of oxygen transport in channel, Eq.\eqref{eq:tZchan}, 
and in the cathode catalyst layer\cite{Kulikovsky_17b} also exhibit negative real part
(Figure~\ref{fig:Zox}). Last but not least, in low--Pt cells, an important role 
plays oxygen transport through a thin 
Nafion film covering Pt/C agglomerates in the cathode catalyst 
layer\cite{Greszler_12,Weber_14,Mathias_16}.
The spectrum of this transport layer is quite similar in shape to the spectrum 
in Figure~\ref{fig:Zgdl}\cite{Kulikovsky_21b}.
Thus, all the oxygen transport processes in a PEMFC cannot be described 
by the standard $RC$--kernel and  another kernel suitable 
for description of impedance elements with the negative real part is needed.  

\begin{table}
\small
\begin{tabular}{|l|c|}
\hline
   GDL thickness $l_b$, cm              & 0.023  \\ 
   Catalyst layer thickness $\lcat$, cm & $10\cdot 10^{-4}$ (10 $\mu$m)       \\
   ORR Tafel slope $b$, mV              &  30 \\
   Double layer capacitance $\Cdl$, F~cm$^{-3}$  &  20   \\ 
   GDL oxygen diffusivity $D_b$, cm$^2$~s$^{-1}$ &  0.01 \\      
   Cell current density $J$, A~cm$^{-2}$  &  0.1   \\
   Pressure                             & Standard   \\
   Cell temperature $T$, K              & 273 + 80   \\ 
   Aur flow stoichiometry $\lam$        & 2.0 \\ 
\hline
\end{tabular}
\caption{The base--case cell parameters used in calculations. 
  }
\label{tab:parms}
\end{table}

In a standard PEMFC, unless the cell current density is small,
the DRT peaks of oxygen transport in the GDL, CCL and channel 
are expected to locate at frequencies below the frequency $f_{ct}$ of faradaic
(charge transfer) processes. Thus, to capture the oxygen transport peaks, 
correction for negative real part of impedance is needed 
in the range of frequencies $f < f_{ct}$. 
The kernel $K_2$ suggested in his work consists, thus, of two parts: 
\begin{equation}
   K_2 = \left\{
   \begin{split}
      &\dfrac{\tanh\sqrt{\ri\omega\tau}}
             {\sqrt{\ri\omega\tau}\left(1 + \ri\omega\tau\right)}, \quad f \leq f_*, \quad \text{TL kernel} \\
      & \dfrac{1}{1 + \ri\omega\tau}, \quad f > f_*, \quad \text{$RC$ kernel}
   \end{split}
   \right.
   \label{eq:K2def}
\end{equation}
where $f_* \simeq f_{ct}$ is the threshold frequency. 
Selection of optimal $f_*$ is discussed below.
A function similar to impedance of 
a transport layer (TL) \eqref{eq:tZgdl} forms the low--frequency part of $K_2$ ; 
the real part of this function is negative at $\omega\tau > 1.81052$. 
The high--frequency ($RC$) part of 
$K_2$ is the standard $RC$--circuit kernel. Switching between TL and $RC$--kernels 
is necessary, as the TL--part itself does not describe well $RC$--circuit 
impedance (see below). The idea behind Eq.\eqref{eq:K2def} is thus  
to expand the low--frequency components 
of cell impedance using the TL--kernel, and the high--frequency components 
using the standard $RC$--kernel.

It is convenient to combine Eq.\eqref{eq:K2def} into one function
\begin{equation}
   K_2 = \dfrac{\tanh\left(\alpha\sqrt{\ri\omega\tau}\right)}
                             {\alpha\sqrt{\ri\omega\tau}\left(1 + \ri\omega\tau\right)}
   \label{eq:K2}             
\end{equation}
where $\alpha$ is a step function of the frequency $f$
\begin{equation}
   \alpha = 1 - H(f - f_*) + \epsilon,
   \label{eq:alpha}
\end{equation}
$H(x)$ is the Heaviside step function and $\epsilon = 10^{-10}$ is a small 
parameter to avoid zero division error. 
Parameter $\alpha$, therefore, changes from 1 to 0 at the threshold frequency $f = f_*$. 
With $\alpha \to 0$, the Warburg factor in Eq.\eqref{eq:K2} tends to unity:
\begin{equation}
   \dfrac{\tanh\left(\alpha\sqrt{\ri\omega\tau}\right)}
                             {\alpha \sqrt{\ri\omega\tau}} \to 1,
    \quad \text{as}\quad \alpha \to 0 
\end{equation} 
and hence the $\alpha$--function serves as a switch between 
the two kernels in Eq.\eqref{eq:K2def}.

With Eq.\eqref{eq:K2},  Eq.\eqref{eq:drtG} takes the form
\begin{equation}
   Z = \Rinf + R_{pol}
               \int_{-\infty}^{\infty}
                  \dfrac{\tanh\left(\alpha\sqrt{\ri\omega\tau}\right)G(\tau)\,d\ln(\tau)}
                        {\alpha\sqrt{\ri\omega\tau}\left(1 + \ri\omega\tau\right)}.
   \label{eq:drtK2}
\end{equation} 
Setting  in Eq.\eqref{eq:drtK2} $\omega \to 0$, it is easy to show 
that $G$ still obeys to normalization conditions Eq.\eqref{eq:normG}.

\section{Numerical results and discussion}

\subsection{Synthetic impedance tests}

\begin{figure}
\begin{center}
\includegraphics[scale=0.33]{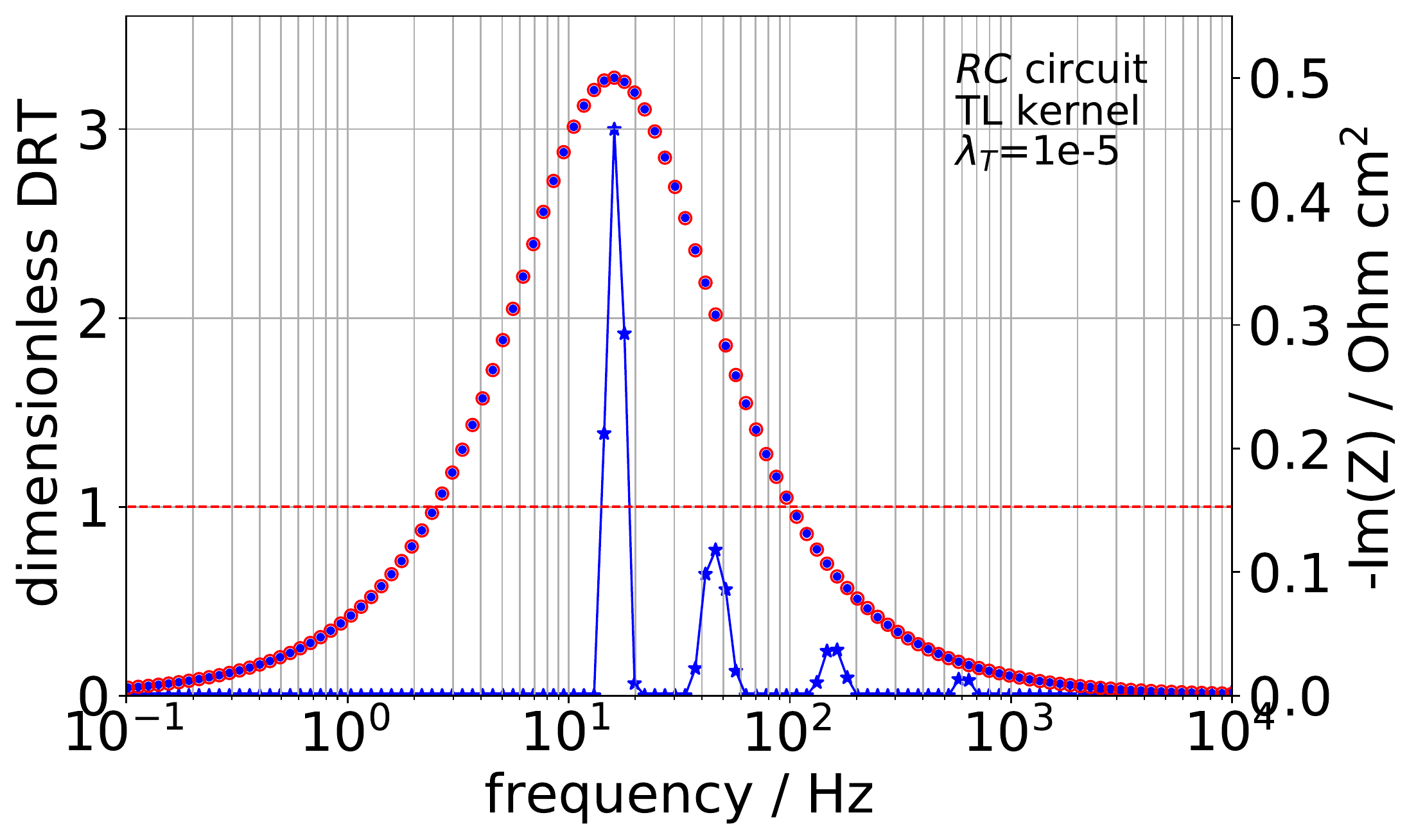}
\caption{DRT (solid line, blue stars) and imaginary part of $RC$--circuit 
   impedance $1 / (1 + 10^{-2}\ri\omega)$ (blue dots)  
   calculated with Eq.\eqref{eq:drtK2} and $\alpha =1$ 
   for all frequencies. Open red circles show imaginary 
   part reconstructed from the calculated DRT using Eq.\eqref{eq:drtK2}.
   Dashed red line -- plot $\alpha=1$ for this calculation.     
  }
\label{fig:Zrc}
\end{center}
\end{figure}

Figure~\ref{fig:Zrc} shows imaginary part of $RC$--circuit impedance for $R=1$, $C=0.01$ 
and the DRT calculated using Eq.\eqref{eq:drtK2} and $\alpha=1$ 
for all frequencies (TL kernel). 
As can be seen, the main peak is positioned correctly 
at the frequency $1/(2\pi R C)$; however, the DRT spectrum 
exhibits three phantom peaks located
to the right of the main peak. It is worth noting that quite similarly, 
the $RC$--kernel generates several phantom peaks in the DRT of Warburg 
finite--length impedance\cite{Kulikovsky_20b}, i.e., no single kernel is able 
to represent correctly DRT of all impedance components. 

In a standard PEMFC, the characteristic frequencies of channel and GDL impedance are about 1 and 10 Hz,
respectively\cite{Kulikovsky_21h}. Thus, the typical value of the threshold 
frequency $f_*$ in Eq.\eqref{eq:alpha} should be about 10 Hz; however, the exact value 
can always be selected simply looking at the calculated DRT spectrum.
In standard PEMFCs operated at oxygen stoichiometry of 2, the faradaic DRT peak is located  
to the right of the GDL transport peak on the log--frequency scale (see below). 

\begin{figure}
\includegraphics[scale=0.31]{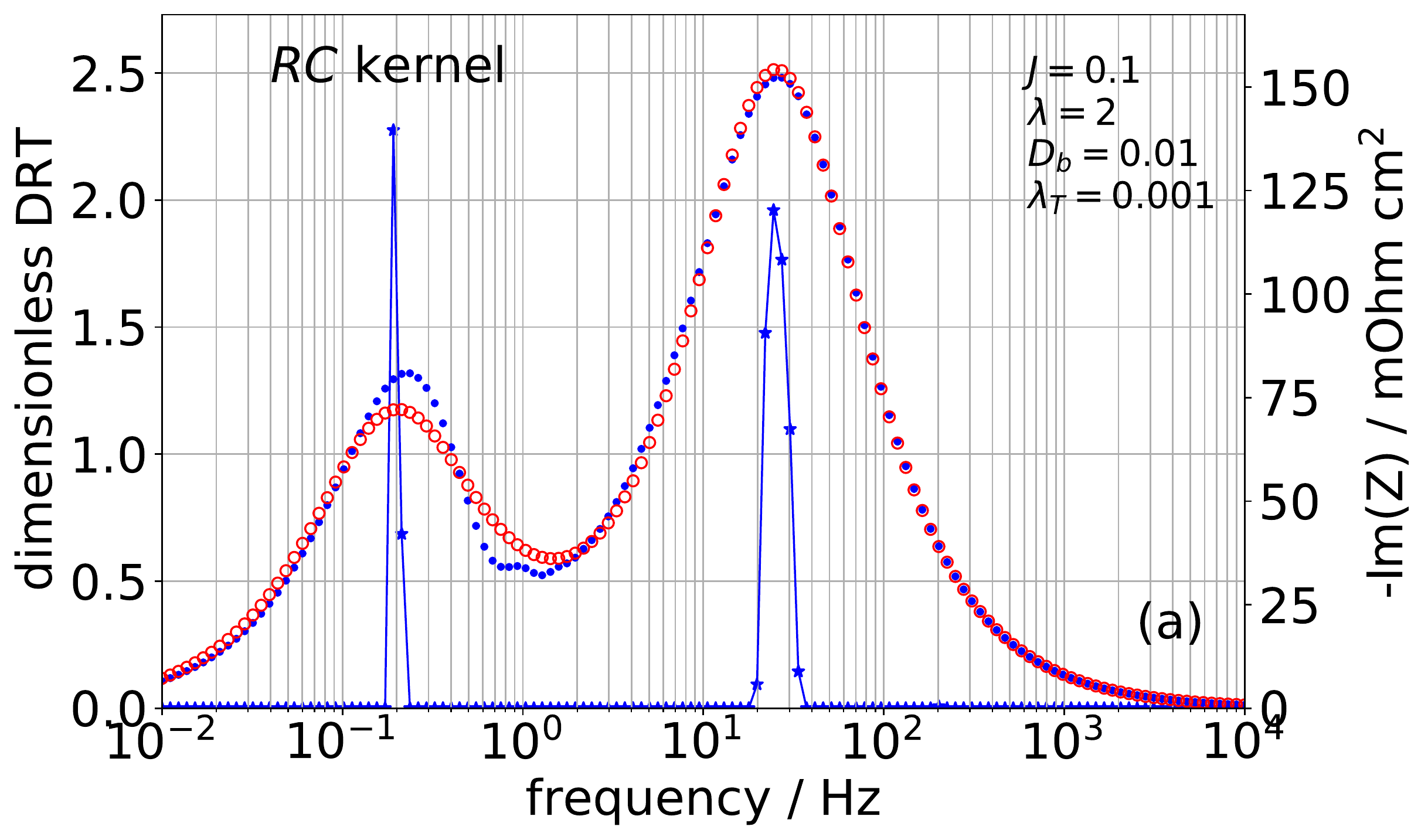}
\includegraphics[scale=0.31]{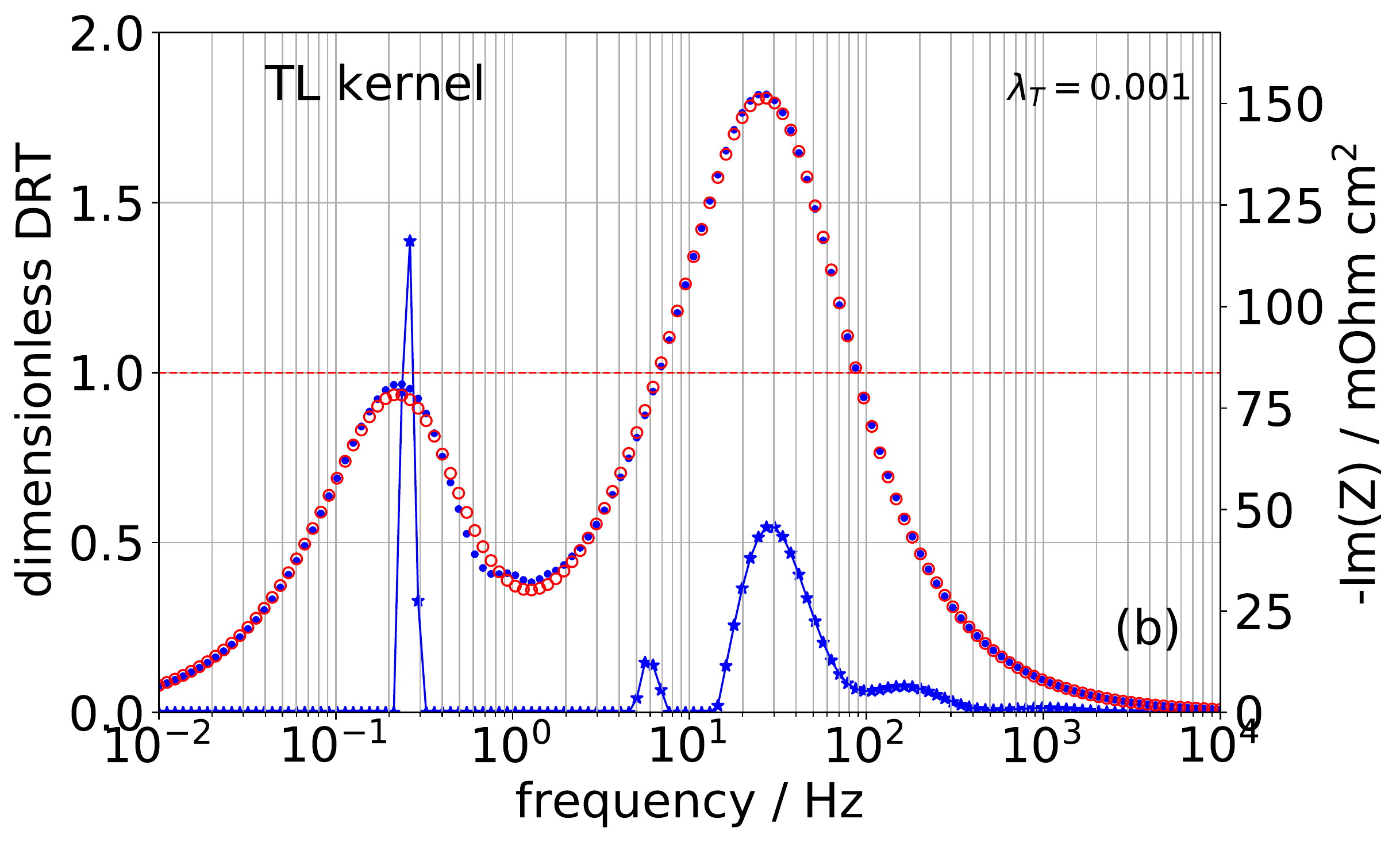}
\includegraphics[scale=0.31]{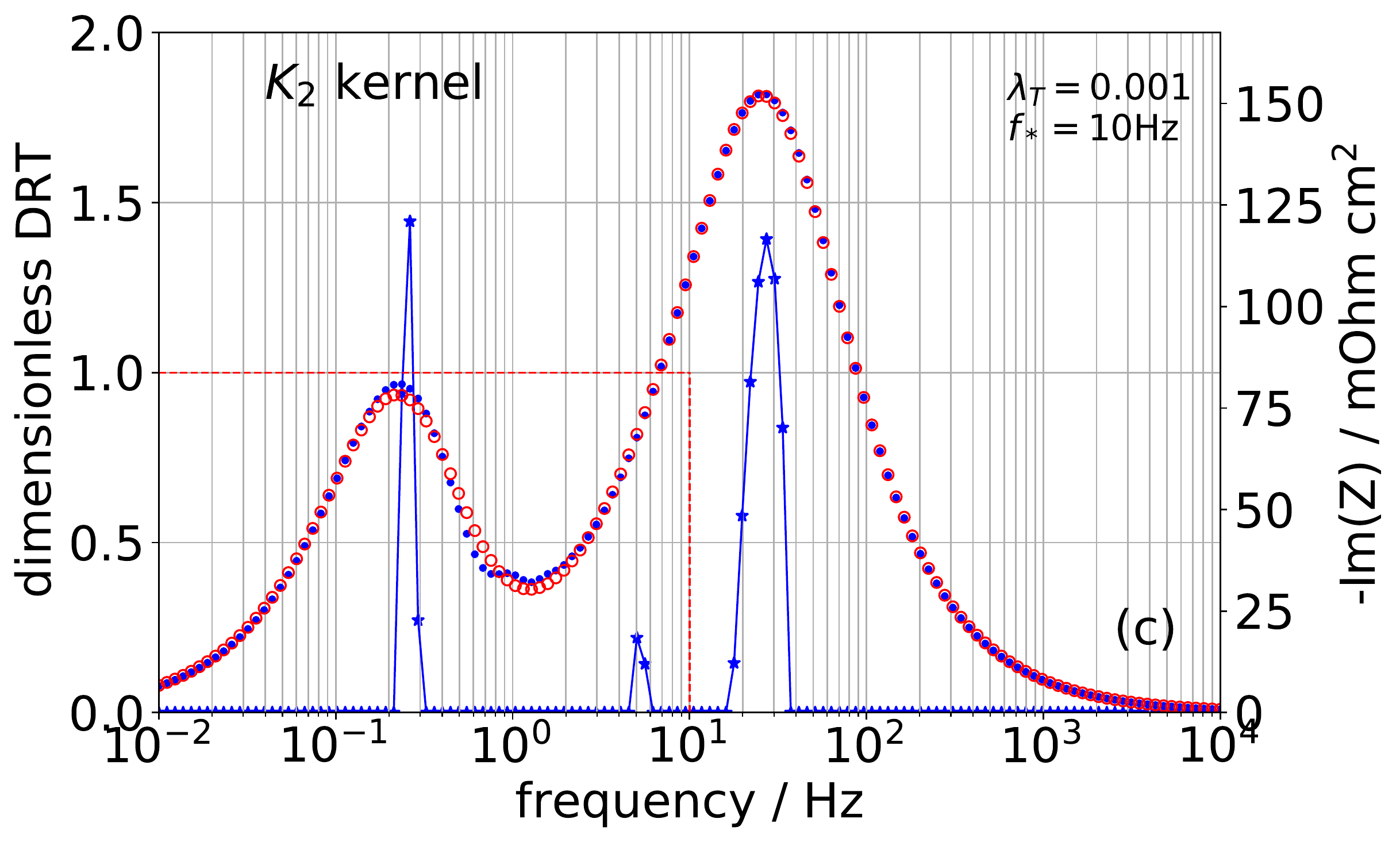}
\hspace*{-3mm}\includegraphics[scale=0.295]{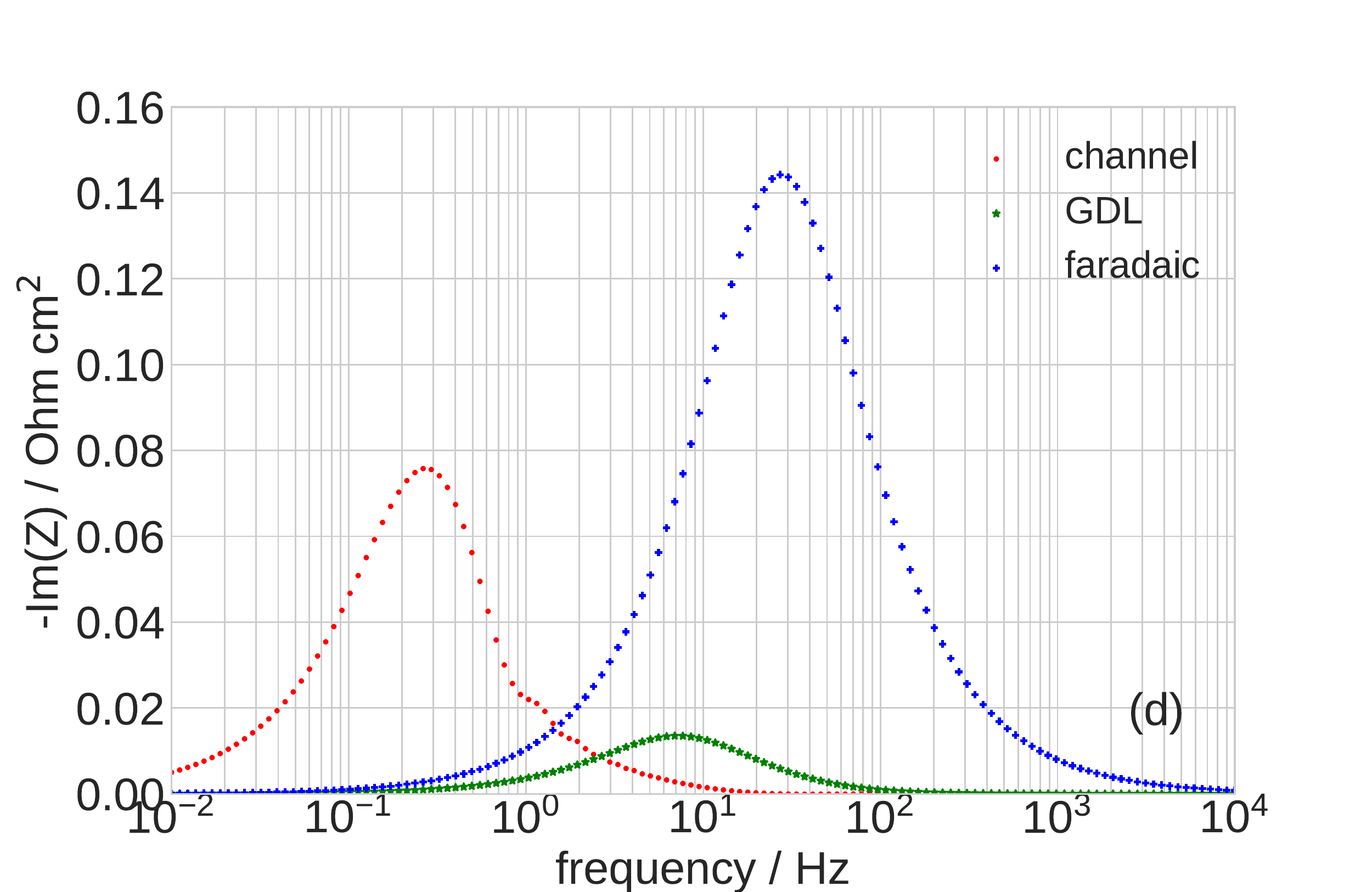}
\caption{ (a) DRT calculated using imaginary part of the total 
   cathode side impedance, Eq.\eqref{eq:tZtot} (blue dots) 
   with the $RC$--kernel (solid line). Red open circles -- $\Im{Z_{tot}}$ reconstructed from the 
   calculated DRT.
   (b) The same as in (a) curves obtained with the $TL$--kernel  ($\alpha=1$) over the whole 
   frequency range. Red dashed line indicates the plot of $\alpha$--function, Eq.\eqref{eq:alpha}.
   (c) The curves as in (a) obtained with the $K_2$--kernel and threshold frequency $f_*=10$ Hz.
   (d) Separate imaginary parts of the channel, GDL and faradaic impedance.        
  }
\label{fig:DRT3}
\end{figure}

Analytical impedance $\tZ_{tot}$ of the PEMFC cathode side 
has been obtained in\cite{Kulikovsky_21h}  assuming fast proton 
and oxygen transport in the CCL. Equation for $\tZ_{tot}$ includes 
three components: impedance $\tZ_{chan}$ due to oxygen transport in channel,
impedance $\tZ_{gdl}$ of oxygen transport in the GDL, and faradaic (charge--transfer)
impedance $\tZ_{ct}$ The respective formulas are listed in Appendix; 
these solutions allow us to check 
how well DRT from Eq.\eqref{eq:drtK2} captures the channel, GDL and faradaic components 
in the total impedance $\tZ_{tot}$ spectrum (Figure~\ref{fig:DRT3}). 

The spectrum of $\tZ_{tot}$ has been calculated in the frequency range 
of $10^{-2}$ to $10^4$ Hz with 22 points per decade. 
Parameters for impedance calculation are listed 
in Table~\ref{tab:parms}. Figure~\ref{fig:DRT3}d depicts imaginary part of the impedance 
components calculated using equations in Appendix. 
Imaginary part of $\tZ_{tot}$ has then been used 
for calculation of DRT using our recent algorithm based on Tikhonov's regularization
and nonnegative least--squares (NNLS) solver\cite{Kulikovsky_20b,Kulikovsky_21b}. 
The NNLS method greatly outperforms  
projected gradient iterations suggested in\cite{Kulikovsky_20b}. 
 In all the cases, the $L$--curve method gave 
the regularization parameter of $10^{-3}$. 


%

Figure~\ref{fig:DRT3}a shows the DRT spectrum of $\tZ_{tot}$, Eq.\eqref{eq:tZtot}, 
calculated using the $RC$--kernel.  As can be seen, the standard kernel returns 
only two peaks corresponding to the channel (left peak) and faradaic
(right peak) impedance. The GDL peak, which is clearly seen 
in Figure~\ref{fig:DRT3}d is completely missing. 
Note poor quality of reconstructed imaginary part (red open circles) between 0.1 and 20 Hz. 
This is a result of poor description of the cell impedance by Eq.\eqref{eq:drtG} 
in the frequency domain where the real part of GDL impedance is negative. 

Figure~\ref{fig:DRT3}b displays the DRT calculated with 
the ``pure'' TL--kernel, Eq.\eqref{eq:K2def}.
The GDL peak is resolved  and the quality of reconstructed imaginary
part is much better; however, phantom high--frequency peaks to the
right of the faradaic peak are clearly seen (cf. Figure~\ref{fig:Zrc}).  
Figure~\ref{fig:DRT3}c shows the DRT spectrum calculated using 
the $K_2$--kernel with the threshold frequency of 10 Hz; 
the GDL peak is well resolved and the phantom peaks vanish. 

Table~\ref{tab:peaks} shows the resistivities, Eq.\eqref{eq:RnG}, 
corresponding to individual peaks in Figure~\ref{fig:DRT3}. As can be seen,   
$K_2$--kernel provides good estimate of the channel and faradaic 
resistivities; however, the GDL resistivity $R_{gdl}$ is underestimated by 30\%.
Nonetheless, as the contribution of $R_{gdl}$ is small, the 30\%--accuracy 
could be tolerated.    

\begin{table}
\small
\begin{tabular}{|l|c|c|c|}
\hline
                &  channel   & GDL    & faradaic  \\
\hline
 Exact          &   0.127    & 0.0250 & 0.300     \\
 RC--kernel     &   0.140    & --     & 0.308     \\
 TL--kernel     &   0.125    & 0.0185 & 0.304$^*$ \\
 $K_2$--kernel  &   0.125    & 0.0171 & 0.305     \\       
\hline
\end{tabular}
\caption{Channel, GDL and faradaic resistivities ($\Omega$~cm$^2$) 
   resulted from DRT, Eq.\eqref{eq:RnG}. Star $^*$ indicates the value
   calculated as a sum of faradaic and all high--frequency peaks 
   in Figure~\ref{fig:DRT3}b.
   The first row shows exact data calculated with Eqs.\eqref{eq:R3}. 
}
\label{tab:peaks}
\end{table}

\subsection{Real PEMFC spectra}

A crucial check for the new kernel is calculation of DRT of a real PEM fuel cell.
Impedance spectra of a standard Pt/C--based PEMFC have been measured in the frequency range 
of 0.1 to about $10^3$ Hz with 11 points per decade. 
The cell geometrical parameters and operating conditions 
are listed in Table~\ref{tab:geom}; note that the air flow stoichiometry 
was 2 in this set of measurements. The impedance points in the frequency range above 
$\simeq 10^3$~Hz have been discarded due to effect of cable inductance. More details 
on experimental setup and measuring procedures can be found in\cite{Kulikovsky_19d}.

\begin{table}
\small
\begin{tabular}{|l|c|}
\hline
   GDL thickness $l_b$, cm              & 0.023  \\ 
   Catalyst layer thickness $\lcat$, cm & $12\cdot 10^{-4}$ (12 $\mu$m)       \\
   Cell active area, cm$^2$             & 76    \\
\hline 
   Cathode pressure, kPa                & 150   \\
   Cathode flow RH                      & 50\%  \\
   Cell temperature $T$, K              & 273 + 80   \\ 
   H$_2$/air flow stoichiometry $\lam$  & 2 / 2      \\ 
\hline
\end{tabular}
\caption{ 
  }
\label{tab:geom}
\end{table}

\begin{figure}
\begin{center}
\includegraphics[scale=0.3]{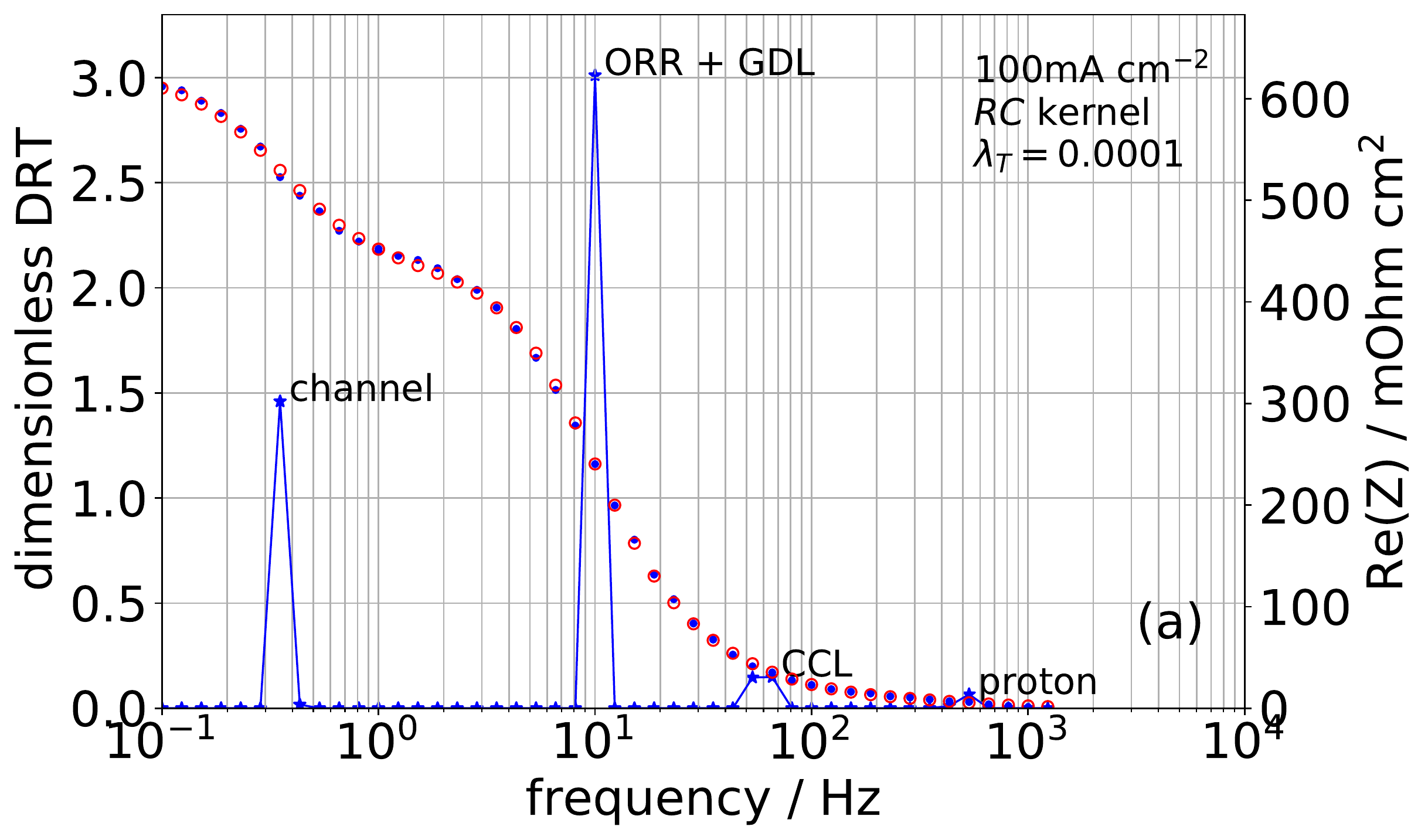}
\includegraphics[scale=0.3]{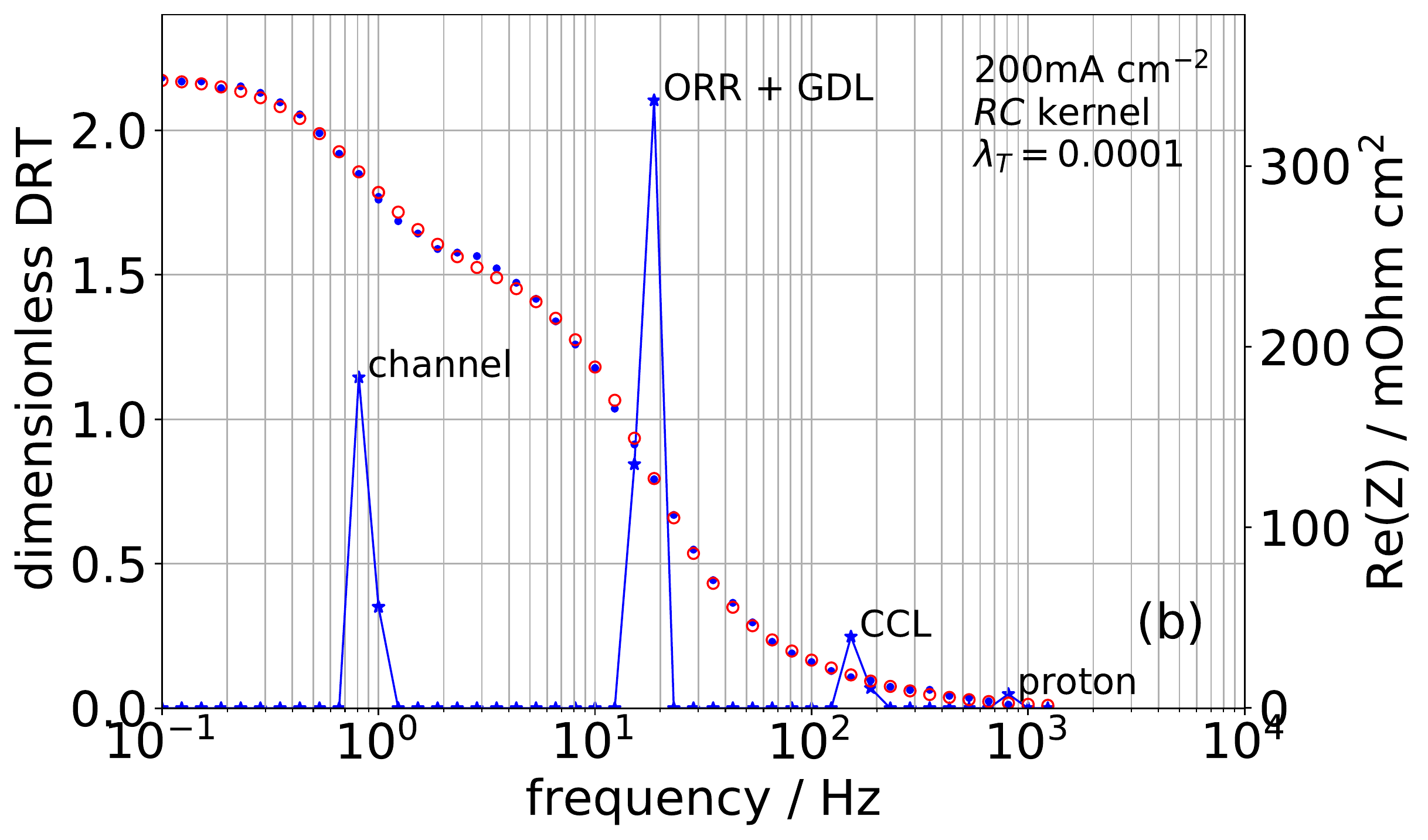}
\includegraphics[scale=0.3]{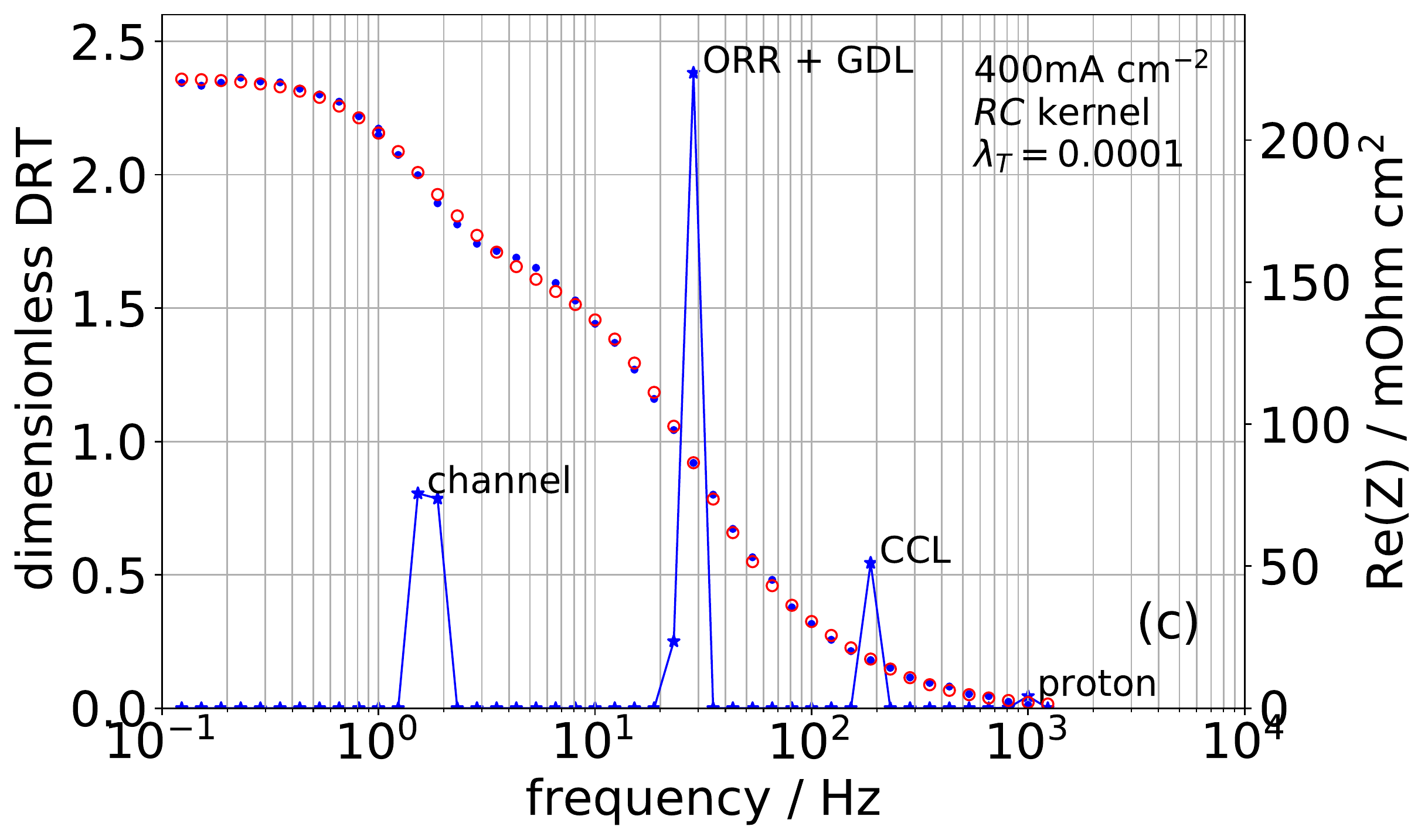}
\includegraphics[scale=0.3]{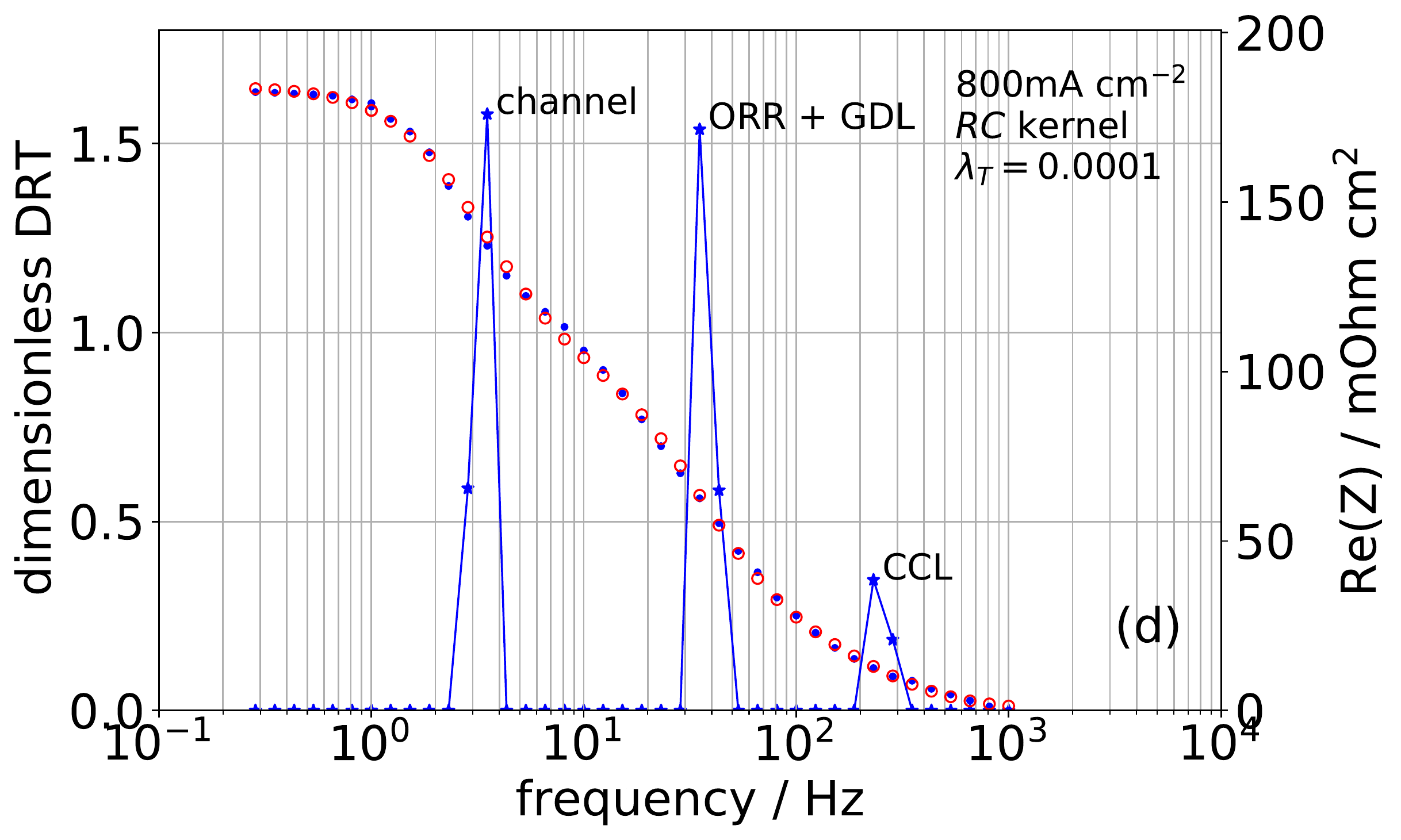}
\caption{DRT (solid line) calculated using real part of the experimental PEMFC
   impedance (blue dots) with the $RC$--kernel
   for the current densities (a) 100, (b) 200, (c) 400, and 
   (d) 800 mA~cm$^{-2}$. Red open circles -- $\Re{Z}$ reconstructed 
   from the calculated DRT.
  }
\label{fig:S067RC}
\end{center}
\end{figure}

\begin{figure}
\begin{center}
\includegraphics[scale=0.55]{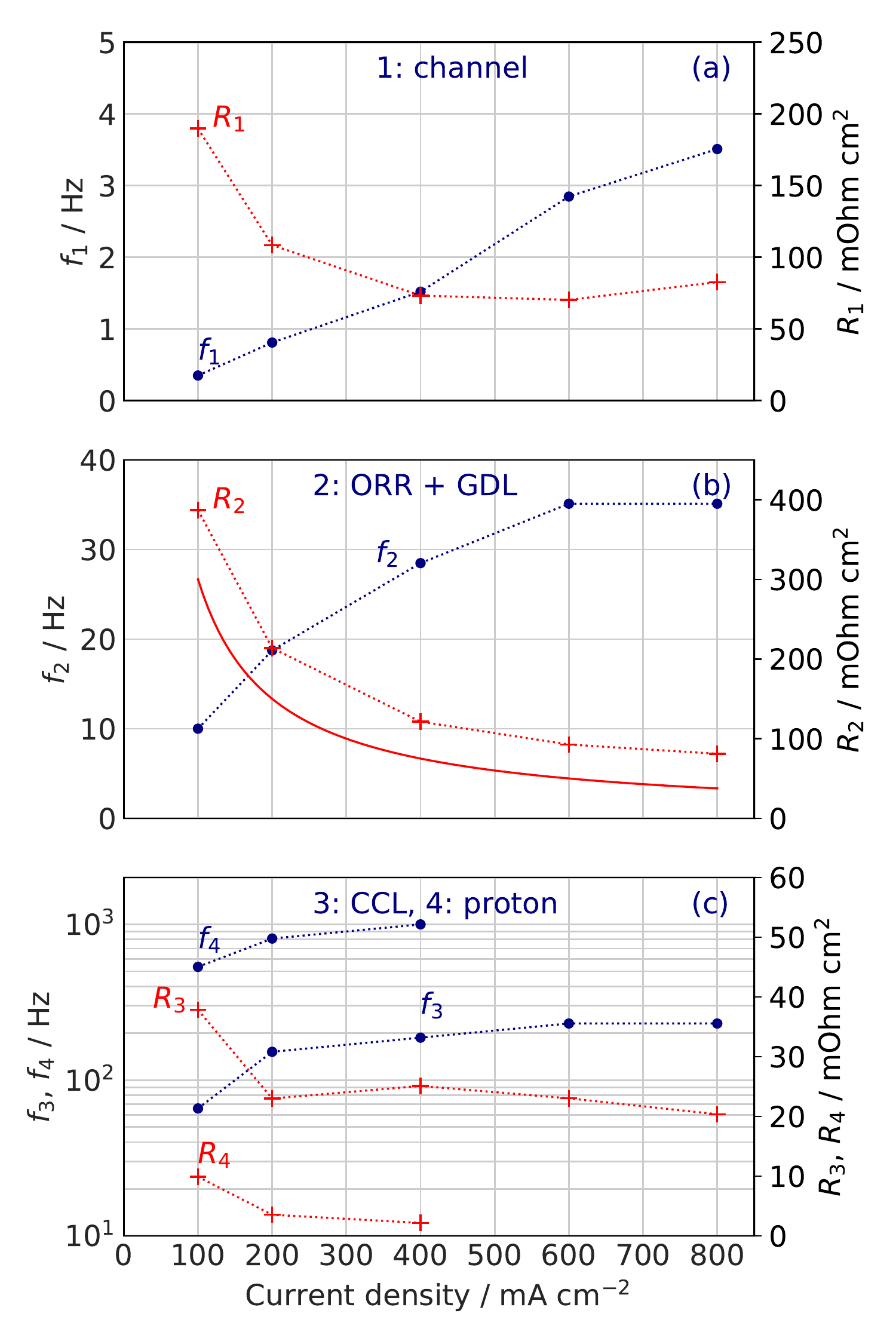}
\caption{Frequency (blue curves, left axis) and resistivity 
   (red curves, right axis) of the DRT peaks 
   calculated using $RC$--kernel (Figure~\ref{fig:S067RC}) 
   vs mean cell current density $J$.  
  }
\label{fig:RCparms}
\end{center}
\end{figure}

Figure~\ref{fig:S067RC} shows DRT spectra calculated 
with the real part of measured impedance using the $RC$--kernel.
Figure~\ref{fig:RCparms} shows the respective peak frequencies 
and resistivities. The DRT spectra in Figures~\ref{fig:S067RC}a--c exhibit four peaks, while 
in Figure~\ref{fig:S067RC}d, the most high--frequency peak disappears.
This peak represents proton transport in the CCL and at high cell currents 
it shifts to frequencies that have been discarded. The characteristic 
frequency $f_4$ of proton transport in the CCL is given by\cite{Kulikovsky_20f}
\begin{equation}
   f_4 \simeq \dfrac{2\sion}{\Cdl \lcat^2}
   \label{eq:fp}
\end{equation}
With the typical values of $\sion \simeq 0.01$~S~cm$^{-1}$ and $\Cdl \simeq 20$~F~cm$^{-3}$ 
(Ref.\cite{Kulikovsky_19d}) we get $f_p \simeq 700$~Hz, which by the order of magnitude 
agrees with the proton peak position in Figures~\ref{fig:S067RC}a--c. The growth 
of $f_4$ with the cell current and the respective decay of the peak resistivity $R_4$  
(Figure~\ref{fig:RCparms}c) is due to growing amount of liquid water improving 
the CCL proton conductivity.  
   
The leftmost peak in Figures~\ref{fig:S067RC}a--d represents impedance due to oxygen 
transport in the cathode channel. The characteristic frequency $f_1$ of this peak linearly
increases with the cell current density (Figure~\ref{fig:RCparms}a), which is a signature 
of channel impedance\cite{Kulikovsky_21h}. 

The highest, second peak in the DRT spectra (Figures~\ref{fig:S067RC}a--d)
represents the contributions of ORR and oxygen transport in the GDL
(see below). In the absence of strong oxygen and proton transport limitations, 
the ORR resistivity is given by
\begin{equation}
   R_{ORR} = \dfrac{b}{J}
   \label{eq:Rorr}
\end{equation}
which follows from the Tafel law.
Qualitatively, the shape of second peak resistivity $R_2$ 
follows the trend of Eq.\eqref{eq:Rorr}
due to dominating contribution of ORR resistivity to this peak (Figure~\ref{fig:RCparms}b).
Note that the separate GDL peak is not resolved by the $RC$--kernel.

The third, CCL--peak in Figures~\ref{fig:S067RC}a--d 
is most probably due to oxygen transport in the CCL pores. For the estimate 
we take the Warburg finite--length formula for the transport layer frequency $f_W$ 
\begin{equation}
   f_W = \dfrac{2.54 D}{2\pi l^2}
   \label{eq:fW}
\end{equation}
where $D$ is the oxygen diffusivity in the transport layer of the thickness $l$.
Setting $f_W = 35$~Hz (Figure~\ref{fig:S067RC}a) and $l=\lcat$, for the 
oxygen diffusion coefficient in the CCL we 
get $\Dox \simeq 1.2\cdot 10^{-4}$~cm$^2$~s$^{-1}$, which agrees with
measurements\cite{Kulikovsky_19d}. With the growth of cell current 
the peak frequency $f_3$ rapidly shifts to 200 Hz (Figure~\ref{fig:RCparms}c), 
corresponding to $\Dox \simeq 7\cdot 10^{-4}$~cm$^2$~s$^{-1}$. The reason for this 
fast growth of $\Dox$ yet is unclear. 

%

\begin{figure}
\begin{center}
\includegraphics[scale=0.3]{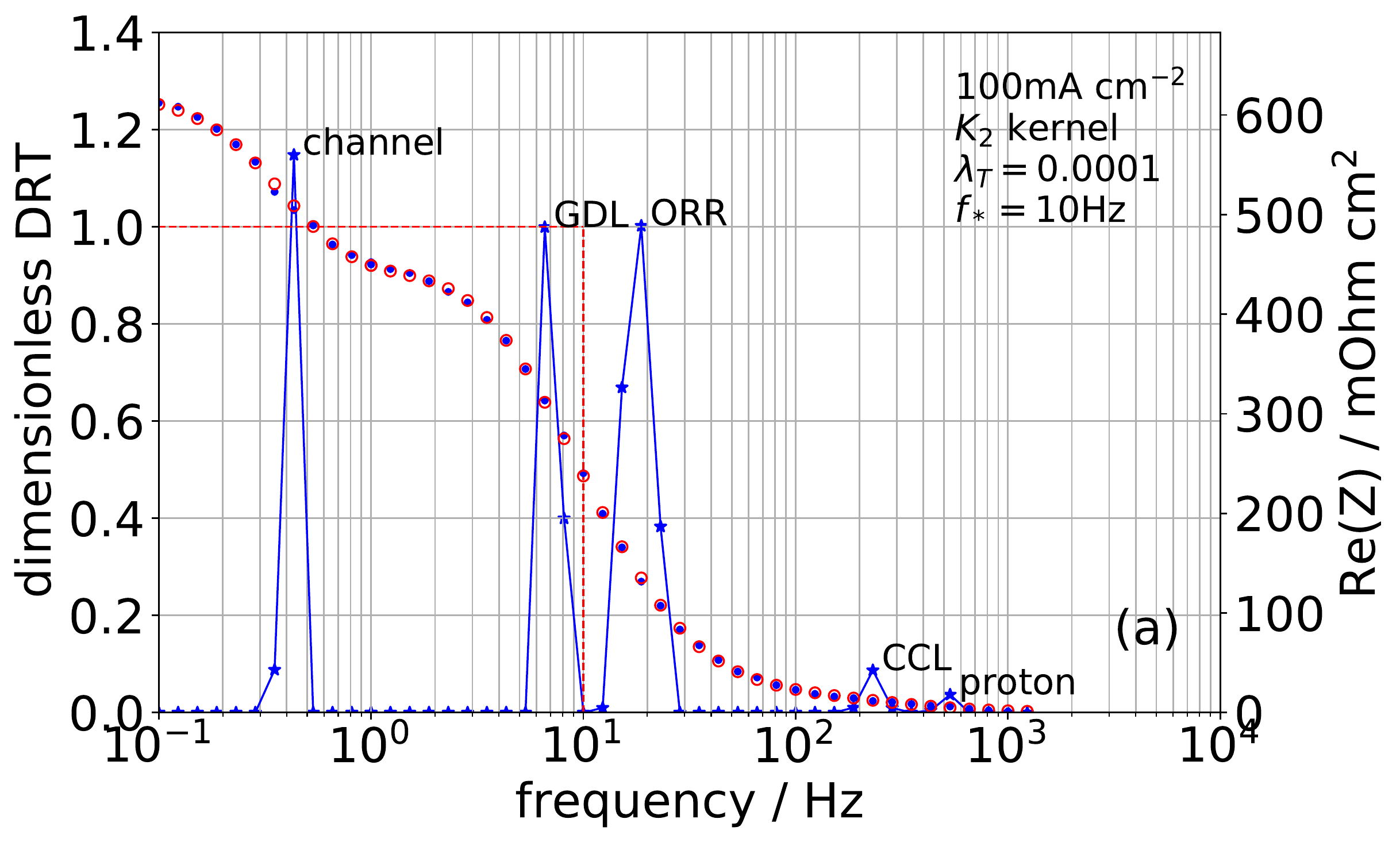}
\includegraphics[scale=0.3]{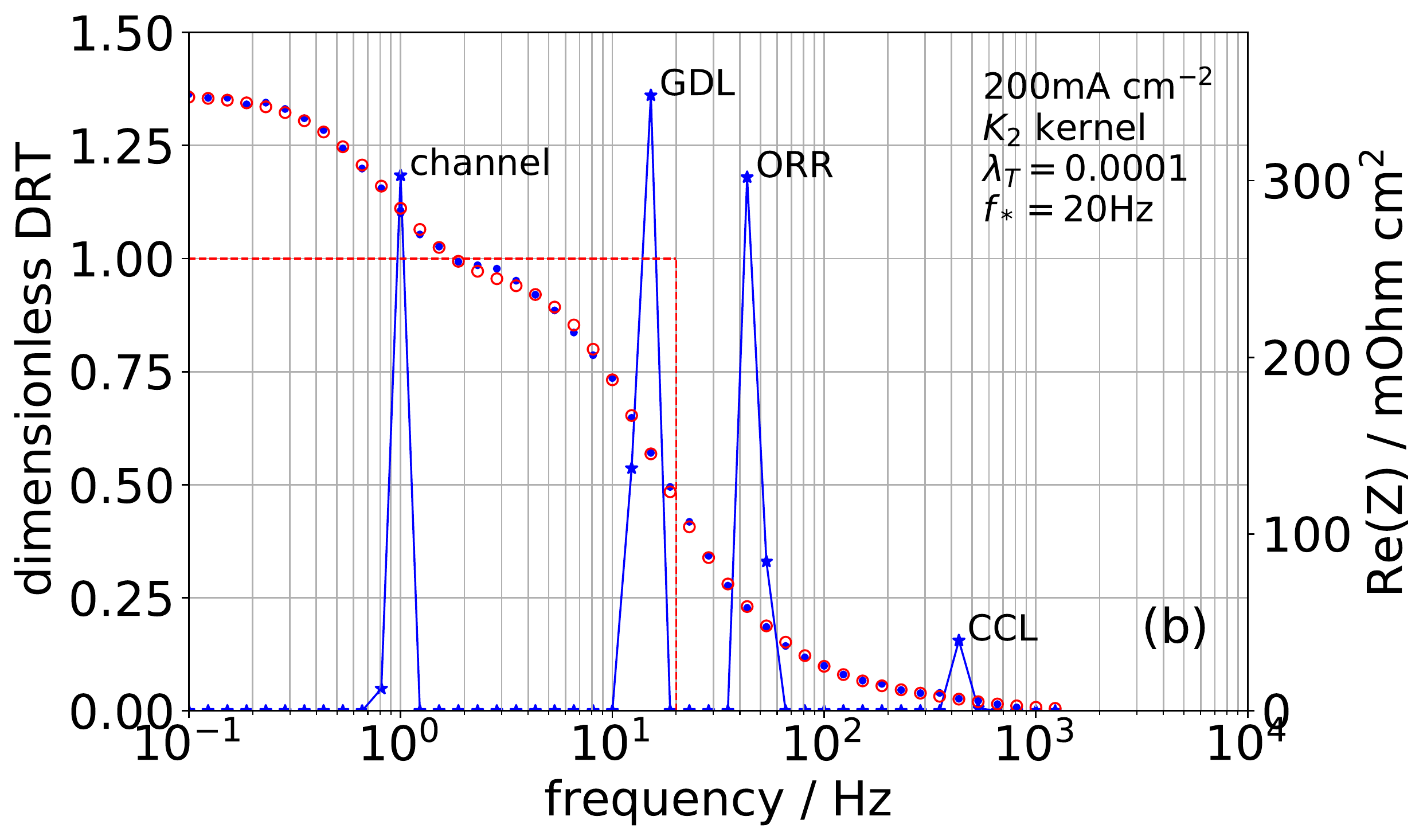}
\includegraphics[scale=0.3]{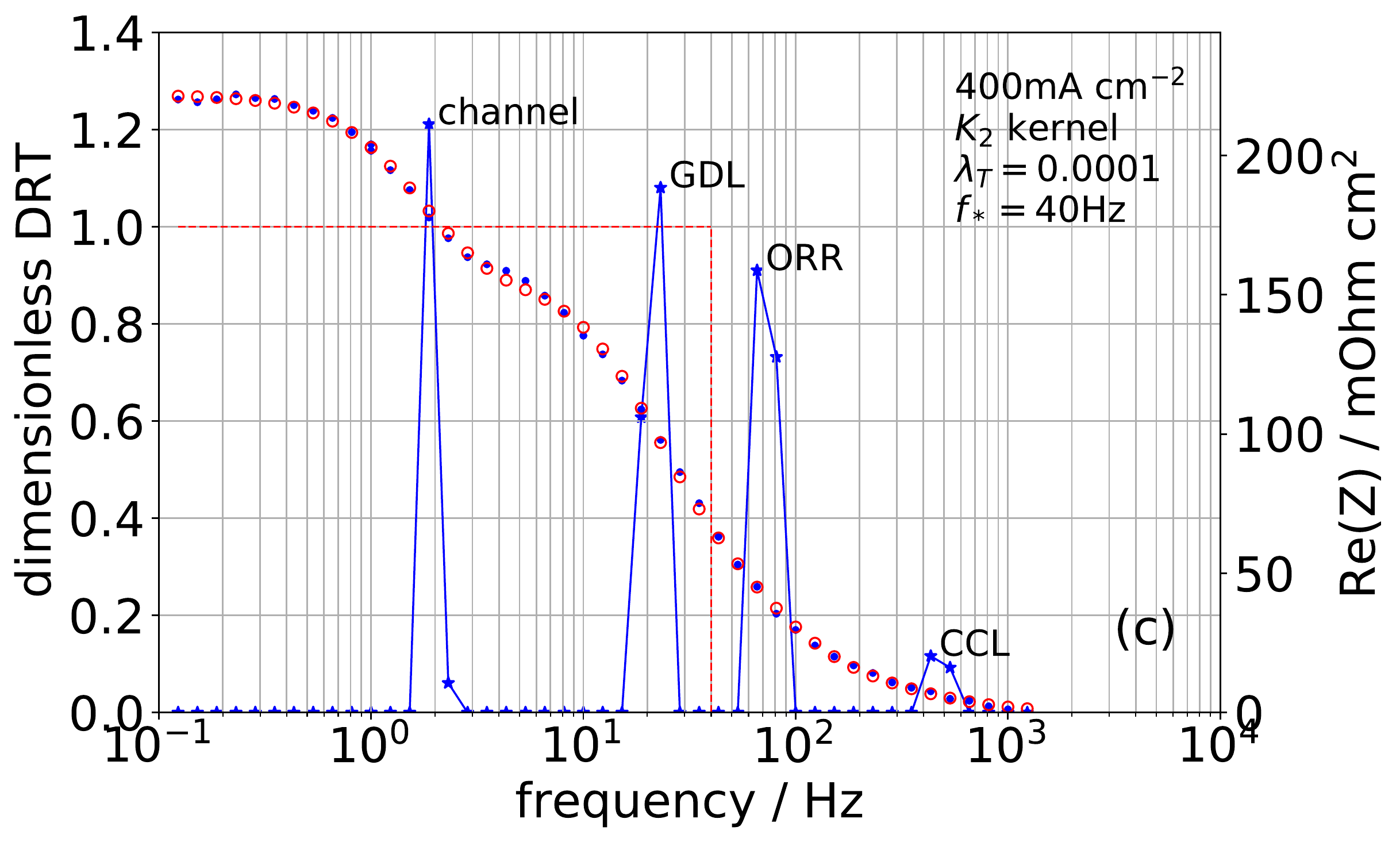}
\includegraphics[scale=0.3]{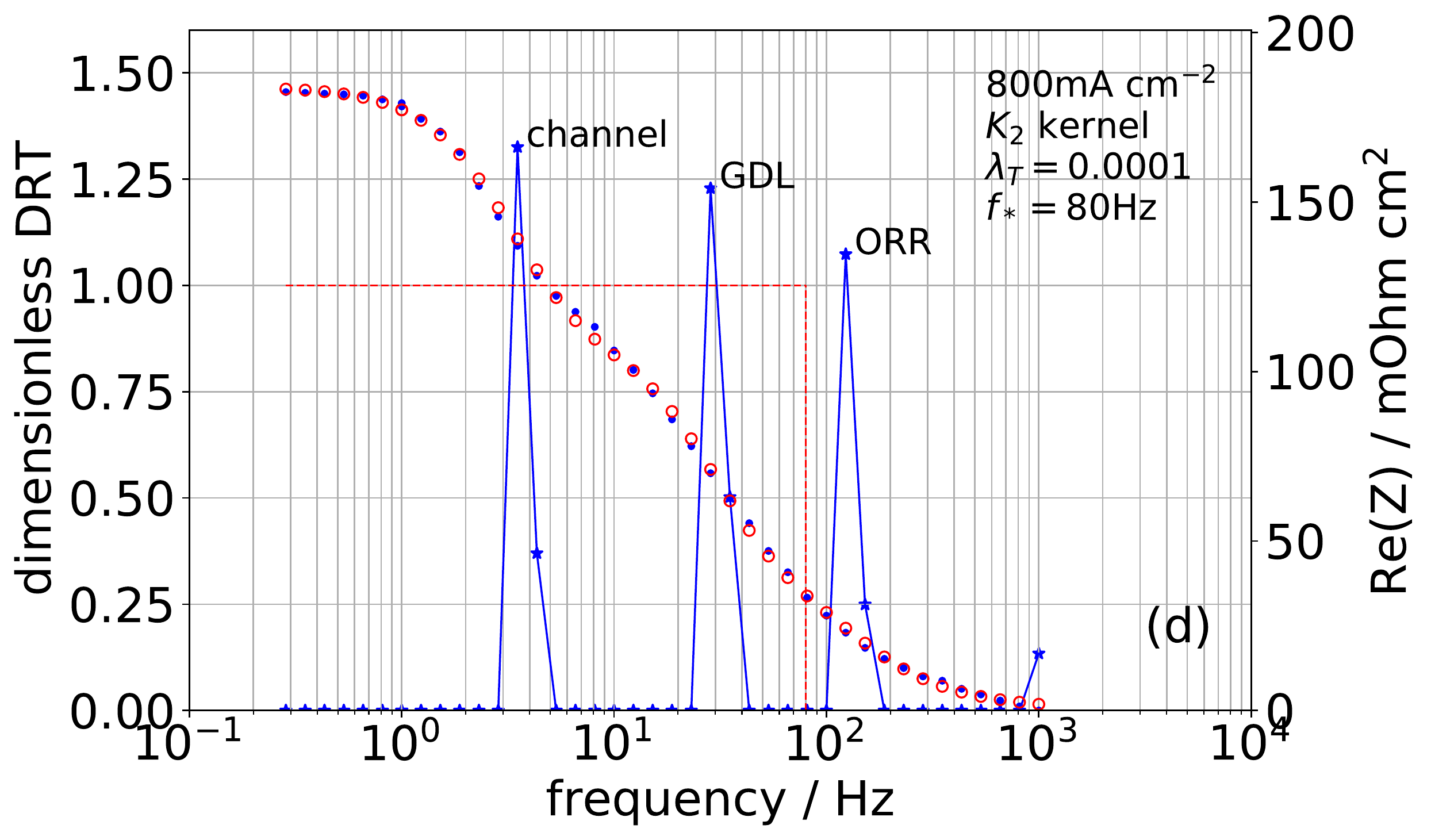}
\caption{ DRT (solid line) calculated using real part of the experimental PEMFC
   impedance (blue dots) with the $K_2$--kernel 
   for the current densities (a) 100, (b) 200, (c) 400, and 
   (d) 800 mA~cm$^{-2}$. 
   Red open circles -- $\Re{Z}$ reconstructed from the 
   calculated DRT. Dashed line shows the plot of 
   $\alpha$, Eq.\eqref{eq:alpha}, switching the $K_2$--kernel from TL-- 
   to $RC$--one at the frequency marked by vertical line.
  }
\label{fig:S067KS}
\end{center}
\end{figure}

Figure~\ref{fig:S067KS} shows the DRT of the same impedance spectra 
calculated with the $K_2$ kernel. The properties of $K_2$ kernel are immediately 
seen: setting of the threshold frequency $f_*$ in Eq.\eqref{eq:alpha}
just to the left of the ``ORR+GDL'' peak in Figures~\ref{fig:S067RC}a--d
splits this peak into two well--resolved peaks (Figures~\ref{fig:S067KS}a--d). 
The left peak of this 
doublet corresponds to the GDL impedance and the right peak to the
ORR impedance. With the $K_2$--kernel, the proton transport peak is seen only 
at the smallest cell current density (Figure~\ref{fig:S067KS}a), while 
at higher currents the peak vanishes indicating its shift to the frequencies
above 1 kHz  (Figures~\ref{fig:S067KS}b--d).

\begin{figure}
\begin{center}
\includegraphics[scale=0.55]{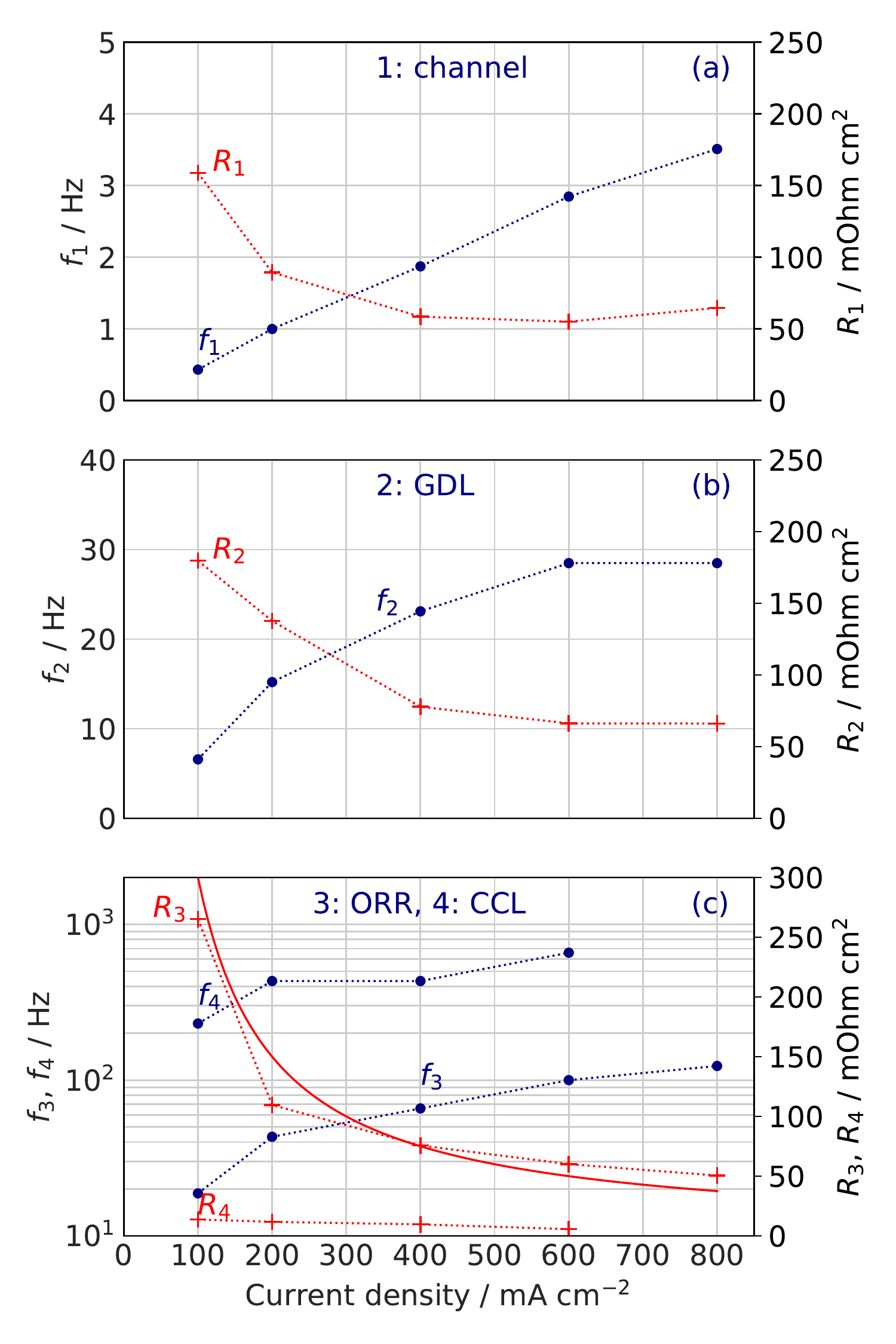}
\caption{Frequency and resistivity of the DRT peaks 
   calculated using $K_2$--kernel (Figure~\ref{fig:S067KS}) 
   vs mean cell current density $J$. Solid line in (c) 
   is the Tafel ORR resistivity $R_{ORR} = b/J$ plotted 
   with the Tafel slope $b=0.03$~V (69 mV/decade).
   }
\label{fig:K2parms}
\end{center}
\end{figure}

The splitting the ``CCL+GDL'' peak into  GDL and ORR peaks 
is confirmed by the behavior of peak resistivities 
in Figures~\ref{fig:K2parms}b and c, respectively. 
The ORR peak resistivity $R_3$ follows the trend of Eq.\eqref{eq:Rorr} 
(solid line in Figure~\ref{fig:K2parms}c)
with the ORR Tafel slope $b=30$~mV, which is a typical value 
for Pt/C cells\cite{Neyerlin_06}. The GDL resistivity $R_2$ decreases 
in the range of cell currents 100 to 400 mA~cm$^{-2}$ and remains 
nearly constant at higher currents (Figure~\ref{fig:K2parms}b). 
Using again the Warburg formula Eq.\eqref{eq:fW}, with $l = l_b = 230\cdot 10^{-4}$ cm
and the frequency between 10 to 30 Hz (Figure~\ref{fig:K2parms}b), for 
the GDL oxygen diffusivity we get quite reasonable 
values of $D_b \simeq 0.013$--0.033~cm$^2$~s$^{-1}$. 
The increase of $D_b$ in the range of 100 to 400 mA~cm$^{-2}$ is probably due 
to growing air flow velocity in the channel at the constant stoichiometry, which 
facilitates liquid droplets removal from the GDL.

$K_2$ kernel returns twice lower resustivity and about twice higher 
frequency of the CCL peak (cf. $f_3$, $R_3$ in Figure~\ref{fig:RCparms}c
and $f_4$, $R_4$ in Figure~\ref{fig:K2parms}c). This shift leads to 
twice higher estimate of the CCL oxygen diffusivity, which is still 
acceptable (see above). Overall, confirmation of the CCL peak nature 
requires measurements at variable oxygen concentration and relative humidity.       


Over the past years, large efforts have been directed toward development 
of universal code capable to calculate DRT based on $RC$--kernel,  
not using any {\em a priori} information on the system
(see a nice review of Effendy, Song and Bazant\cite{Effendy_20}). However, in PEMFC studies 
it would be wasteful to ignore analytical results showing that the $RC$--kernel
alone is not well suited for DRT description of the spectra.

\section{Conclusions}

Impedance of all oxygen transport processes in a PEM fuel cell exhibits 
negative real part in some frequency range. This makes
it difficult accurate calculation of the respective DRT peaks using 
the standard $RC$--kernel $1 / (1 + \ri\omega\tau)$. A novel kernel 
$K_2$, Eq.\eqref{eq:K2}, is suggested. $K_2$ combines the low--frequency 
transport layer kernel having a domain with negative real part, and the standard
$RC$--kernel for description of faradaic and high--frequency processes 
in the cell. Calculation of DRT for analytical PEMFC impedance 
shows that $K_2$ kernel captures 
the peak due to oxygen transport in the gas--diffusion layer, while 
the $RC$--kernel can miss this peak. Comparison of Pt/C PEMFC 
DRT calculated using $RC$-- and $K_2$--kernel shows that the $K_2$--kernel 
resolves the GDL oxygen transport peak, which otherwise is merged 
to the ORR peak when using the standard $RC$--kernel. Overall, the $K_2$--spectra 
of a standard Pt/C PEMFC operating at the air flow stoichiometry $\lam=2$ 
consist of five peaks. In the frequency ascending order, these peaks are 
due to (1) oxygen transport in channel, 
(2) oxygen transport in the GDL, (3) faradaic reactions, 
(4) oxygen transport in the CCL, and (5) proton transport in the CCL. 
If the CCL proton conductivity is high, the peak (5) shifts to the frequencies 
well above 1 kHz, and it may not be resolved due to inductance of measuring 
system.

\section*{Acknowledgments}

The author is grateful to Dr. Tatyana Reshetenko (University of Hawaii) 
for experimental spectra used in this work and useful discussions.

\appendix 
   
\section{Model equations for GDL, channel and faradaic impedance}

Equations of this Section have been derived in\cite{Kulikovsky_21h}. 
\begin{itemize}

\item Channel impedance is

\begin{equation}
   \tZ_{chan} = - \dfrac{4\lam \expo N_c }{\bigl(2\lam\ri\tom + (2\lam - 1)\expo\bigr) D_c}
   \label{eq:tZchan}
\end{equation}
where
\begin{multline}
N_c  = \lam^2\tJ \left(\expot + \left(\tJ + \ri\tom(1 + \xi^2)\right)\times\expo - \xi^2\tom^{2}\right) \\
     \times\lexp{\dfrac{ - \expo - \ri \tom\xi^2}{\lam  \tJ}} + \lam\left(\expo - \lam  \tJ + \ri\tom\xi^2\right) \\
     \times\left(\expot + \left(\tJ + \ri\tom(\xi^2 + 1)\right)\expo - \xi^2 \tom^2\right) \\
   - \left( \expo + \ri\tom\xi^2\right)^{2}\expo/2
   \label{eq:Nc}
\end{multline}

\begin{multline}
   D_c = 2\lam^2\tJ\left(\expot + \left(\tJ + \ri\tom(1 + \xi^2)\right)\expo  - \xi^2\tom^2\right) \\
          \times\lexp{\dfrac{ - \expo - \ri \tom\xi^2}{\lam  \tJ}}\expo
   - 2 \lam\xi^6\tom^{4} \\
   + \ri\left(2 \lam\xi^2 - \xi^2 + 4 \lam \right) \xi^{4} \expo \,\tom^{3} \\
      + 2 \xi^2 \left(\left(2 \lam\xi^2 - \xi^2 + \lam \right) \expo + \lam^2\tJ\right)\expo\tom^{2} \\
   - \ri\left(\left(2 \lam - 1\right)\xi^2\expo + 2\lam\tJ\left(\lam\xi^2 - \xi^2 + \lam\right)\right)\expot \tom \\
   - 2\lam\tJ\left(\left(\lam - 1\right)\expo + \lam\tJ\right)\expot.
   \label{eq:Dc}
\end{multline}
and parameters $\xi$ and $\lam$ are given by
\begin{equation}
    \xi = \sqrt{\dfrac{4 F h \cref}{\Cdl \lcat b}},\quad \lam = \dfrac{4 F h v \cref}{L J}.
    \label{eq:xilam}
\end{equation}

\item GDL impedance is given by Eq.\eqref{eq:tZgdl}.

\item Faradaic impedance is 
\begin{equation}
   \tZ_{ct} = \dfrac{1}{\ri\tom + \left(1 - \dfrac{1}{2\lam}\right)\expo}
   \label{eq:tZct}
\end{equation}

\item Total impedance of the cathode side, including channel, GDL and faradaic components
\begin{equation}
   \tZ_{tot} = \dfrac{\lam B^3}{D_{tot}}\left(\cosh(\phi) + \dfrac{\expo\sinh(\phi)}{\psi}\right)
   \label{eq:tZtot}
\end{equation} 
where
\begin{multline}
   D_{tot} = \lam^2\tJ\expo\left(\lam\tJ A + B C\right)\left(\lexp{\dfrac{B}{\lam\tJ}} - 1 \right) \\
                    + \left((\ri\tom + \expo)\lam - \expo/2\right) B^3\cosh(\phi) \\
              - \lam\expo B \left(\lam\tJ A + B(A/2 + C)\right)
   \label{eq:Dtot}
\end{multline}
and the coefficients $A$, $B$ and $C$ are given by 
\begin{equation}
   A = \dfrac{\psi\expo}{\lam\cosh(\phi)\left(\psi + \expo\tanh(\phi)\right)}
   \label{eq:A}
\end{equation}
\begin{equation}
   B = - \ri\tom\xi^2 - \psi\tanh(\phi) - \dfrac{\lam A}{\cosh(\phi)}
   \label{eq:B}
\end{equation}
\begin{equation}
   C = -\dfrac{\psi\left(\ri\tom + \expo\right)}
              {\cosh(\phi)\left(\psi + \expo\tanh(\phi)\right)}
   \label{eq:C}
\end{equation}
Auxiliary parameters $\phi$ and $\xi$ are given by
\begin{equation}
   \begin{split}
      &\phi = \mu\tl_b\sqrt{\sqa} \\
      &\psi = \mu\sqrt{\ri\tom\tD_b} 
   \end{split}
   \label{eq:phipsi}
\end{equation}

\item The cell polarization curve is 
\begin{equation}
   \expo = - \lam\lnl{1 - \dfrac{1}{\lam}} \tJ
   \label{eq:tvcc}
\end{equation}

\item GDL $R_{gdl}$, faradaic $R_{f}$ and channel $R_{chan}$ 
resistivities in the dimension form ($\Omega$~cm$^2$):
\begin{equation}
   \begin{split}
      & R_{gdl} = \dfrac{b l_b}{4 F D_b \cref}, \\ 
      & R_f = \dfrac{b}{J}   \\
      & R_{chan} = - \dfrac{b}{J\,(2\lam -1)}\biggl(2 \lam^2\lnl{1 - \dfrac{1}{\lam}}^2 \\
      & \qquad       - \lam\lnl{1 - \dfrac{1}{\lam}} - \dfrac{2}{\ln(1 - 1/\lam)} \biggr).
   \end{split}
   \label{eq:R3}
\end{equation}

\end{itemize}



\vspace*{2em} 

\centerline{\Large\bf Nomenclature}

\small

\begin{tabular}{ll}
	$\tilde{}$   &  Marks dimensionless variables                             \\
	$b$          &  ORR Tafel slope, V                                        \\
    $\Cdl$       &  Double layer volumetric capacitance, F~cm$^{-3}$            \\
	$c_1$        &  Oxygen molar concentration  \\ 
                 & at the CCL/GDL interface, mol~cm$^{-3}$      \\
    $c_b$        &  Oxygen molar concentration in the GDL, mol~cm$^{-3}$      \\
    $c_h$        &  Oxygen molar concentration in the channel, mol~cm$^{-3}$      \\
	$\cref$      &  Reference (inlet) oxygen concentration, mol~cm$^{-3}$      \\
    $D_b$        &  Oxygen diffusion coefficient in the GDL, cm$^2$~s$^{-1}$    \\  
	$F$          &  Faraday constant, C~mol$^{-1}$                            \\
    $f$          &  Characteristic frequency, Hz                              \\
  	$i_*$        &  ORR volumetric exchange current density, A~cm$^{-3}$          \\
    $\ri$        &  Imaginary unit                                            \\
	$j$          &  Local proton current density along the CCL,  A~cm$^{-2}$     \\
    $\jlim$      &  Limiting current density  \\
                 &  due to oxygen transport in the GDL, Eq.\eqref{eq:jlim} A~cm$^{-2}$     \\
	$j_0$        &  Local cell current density, A~cm$^{-2}$                   \\
    $l_b$        &  GDL thickness,  cm                                       \\
	$\lcat$      &  CCL thickness,  cm                                       \\
    $t$          &  Time, s                                                   \\
    $t_*$        &  Characteristic time, s, Eq.\eqref{eq:tast}                \\
	$x$          &  Coordinate through the cell, cm                           \\
    $Z$          &  Local impedance, Ohm~cm$^2$                               \\
    $Z_{gdlc}$   &  GDL+channel impedance, Ohm~cm$^2$                      \\
    $Z_{tot}$    &  Total cathode side impedance, including \\ 
                 &  faradaic one, Ohm~cm$^2$ \\  
    $z$          &  Coordinate along the cathode channel, cm  \\[1em]              
\end{tabular}

{\bf Subscripts:\\}

\begin{tabular}{ll}
	$0$      & Membrane/CCL interface \\
	$1$      & CCL/GDL  interface     \\
    $b$      & In the GDL \\
    $gdl$    & GDL   \\
    $gdlc$   & GDL+channel \\
    $f$      & faradaic     \\
    $h$      & Air channel             \\
    $W$      & Warburg \\[1em]
\end{tabular}

{\bf Superscripts:\\}

\begin{tabular}{ll}
	$0$      & Steady--state value \\
	$1$      & Small--amplitude perturbation \\[1em]
\end{tabular}

{\bf Greek:\\}

\begin{tabular}{ll}
    $\eta$              &  ORR overpotential, positive by convention, V     \\
    $\lam$              &  Air flow stoichiometry, Eq.\eqref{eq:xilam}      \\
    $\lam_T$            &  Tikhonov regularization parameter \\ 
    $\mu$               &  Dimensionless parameter, Eq.\eqref{eq:mu}        \\
    $\xi$               &  Dimensionless parameter, Eq.\eqref{eq:xilam}      \\
    $\phi$              &  Dimensionless parameter. Eq.\eqref{eq:phipsi} \\
    $\psi$              &  Dimensionless parameter. Eq.\eqref{eq:phipsi} \\
    $\omega$            &  Angular frequency of the AC signal, s$^{-1}$
\end{tabular}

\end{document}